\documentclass[reprint,prb,amsmath,superscriptaddress]{revtex4-2}

\setcitestyle{super}
\usepackage{times}
\usepackage{graphicx}
\usepackage{amssymb}
\usepackage{lineno}
\usepackage{xcolor}
\usepackage{hyperref}

\usepackage[]{changes}

\newcommand{\rfig}[2]{Fig.~\ref{#1}\textbf{#2}}
\newcommand{\rfigs}[2]{Figs.~\ref{#1}\textbf{#2}}

\newcommand\Thot{T_\mathrm{hot}}

\def\rm{\mathrm}

\newcommand{\siqse}{Shenzhen Institute for Quantum Science and Engineering, Southern University of Science and Technology, Shenzhen 518055, China}
\newcommand{\gdpkl}{Guangdong Provincial Key Laboratory of Quantum Science and Engineering, Southern University of Science and Technology, Shenzhen 518055, China}
\newcommand{\iqac}{International Quantum Academy, Shenzhen 518048, China}
\newcommand{\nxy}{School of Physics, Ningxia University, Yinchuan 750021, China}
\newcommand{\hefeilab}{Shenzhen Branch, Hefei National Laboratory, Shenzhen 518048, China}

\begin{document}

\title{A thermal-noise-resilient microwave quantum network traversing 4~K}

\author{Jiawei Qiu}
\thanks{These authors contributed equally: Jiawei Qiu, Zihao Zhang}
\address{\iqac}
\address{\siqse}
\address{\gdpkl}

\author{Zihao Zhang}
\thanks{These authors contributed equally: Jiawei Qiu, Zihao Zhang}
\address{\iqac}
\address{\siqse}
\address{\gdpkl}

\author{Zilin Wang}
\address{\nxy}
\address{\iqac}

\author{Libo Zhang}
\address{\iqac}
\address{\siqse}
\address{\gdpkl}

\author{Yuxuan Zhou}
\address{\iqac}

\author{Xuandong Sun}
\address{\iqac}
\address{\siqse}
\address{\gdpkl}

\author{Jiawei Zhang}
\address{\iqac}
\address{\siqse}
\address{\gdpkl}

\author{Xiayu Linpeng}
\address{\iqac}

\author{Song Liu}
\address{\iqac}
\address{\siqse}
\address{\gdpkl}
\address{\hefeilab}

\author{Jingjing Niu}
\email{niujj@iqasz.cn}  % must be next to \author
\address{\iqac}
\address{\hefeilab}

\author{Youpeng Zhong}
\email{zhongyoupeng@iqasz.cn}
\address{\iqac}
\address{\hefeilab}

\author{Dapeng Yu}
\email{yudapeng@iqasz.cn}
\address{\iqac}
\address{\hefeilab}

\maketitle

% \begin{abstract}
\textbf{Quantum communication at microwave frequencies has been fundamentally constrained by the susceptibility of microwave photons to thermal noise, hindering their application in scalable quantum networks. 
Here we demonstrate a thermal-noise-resilient microwave quantum network that establishes coherent coupling between two superconducting qubits through a 4~K thermalized niobium-titanium transmission line. 
By overcoupling the communication channel to a cold load at 10~mK, we suppress the effective thermal occupancy of the channel to 0.06 photons through radiative cooling -- a two-order-of-magnitude reduction below ambient thermal noise. We then decouple the cold load and rapidly transfer microwave quantum states through the channel while it rethermalizes, achieving a 58.5\% state transfer fidelity and a 52.3\% Bell entanglement fidelity, both exceeding the classical communication threshold.
Our architecture overcomes the temperature-compatibility barrier for microwave quantum systems, providing a scalable framework for distributed quantum computing and enabling hybrid quantum networks with higher-temperature semiconductor or photonic platforms.}
% \end{abstract}

%\vspace{5pt}

\section*{Introduction}
Microwave technology forms the backbone of modern telecommunications, from global satellite links to ubiquitous mobile networks. However, its transition to the quantum domain has been fundamentally limited by the extreme thermal sensitivity of microwave photons--a consequence of their low energy ($\sim$20 $\mu$eV at 5~GHz) that renders single-photon quantum states vulnerable to ambient thermal noise~\cite{Krantz2019}. While Josephson junction based superconducting quantum circuits have revolutionized microwave-frequency quantum information processing~\cite{Blais2021,Devoret2013,Barends2014,Arute2019}, achieving record-breaking qubit counts and gate fidelities~\cite{ZJU2024,Google2024,IBM2024,Gao2024}, their operation remains constrained to millikelvin temperatures within dilution refrigerators (DRs). This creates a critical scalability barrier: monolithic integration of thousands of qubits faces fabrication challenges~\cite{Bravyi2022}, while distributed quantum networks require cryogenically incompatible thermal environments across nodes~\cite{Jiang2007,Monroe2014}.

Recent advances in superconducting quantum links~\cite{Kurpiers2018,Axline2018,Campagne2018,Leung2019,Magnard2020,Storz2023,Yam2023,Qiu2024} and modular architectures~\cite{Zhong2021,Burkhart2021,Niu2023,kannan2023,Mollenhauer2024,Song2024,almanakly2024,deng2025} highlight the potential of microwave quantum networks. However, these systems demand passive ground-state cooling of entire transmission channels, as even the presence of half a photon's worth of thermal noise within the communication channel can effectively obliterate the transmitted quantum signal~\cite{Habraken2012}. Theoretical proposals suggest thermal resilience through dynamic error suppression~\cite{Xiang2017,Vermersch2017}, yet experimental realizations have remained elusive due to the lack of techniques for rapid thermal noise mitigation in open quantum systems.

Here we address this challenge by developing a radiatively cooled microwave quantum network that dynamically modulates channel thermalization. Our architecture leverages three key innovations: (1) a tunable coupler achieving two-order-of-magnitude on/off ratio between a 4~K transmission line and 10~mK cold load, (2) radiative cooling~\cite{Raman2014,Albanese2020,Xu2020,Wang2021} via strategic over-coupling to suppress thermal occupancy of the channel to 0.06 photons, and (3) time-domain thermal management enabling high-speed quantum state transfer during channel rethermalization. This approach transforms thermal noise from an insurmountable obstacle into a dynamically manageable resource, achieving 58.5\% quantum state transfer fidelity and 52.3\% Bell-state entanglement fidelity--both surpassing the classical communication threshold.
The demonstrated thermal resilience stems from bridging the quantum engineering of superconducting circuits and non-equilibrium thermodynamics. By decoupling the qubit operating temperature from that of the communication channel, our work enables hybrid quantum systems that integrate superconducting processors with various platforms operating at elevated temperatures~\cite{Xiang2013,clerk2020,Huang2024a}. More broadly, it establishes radiative cooling as a generic tool for thermal management in microwave quantum technologies, from distributed quantum computing to hybrid quantum systems.

%%%%% experiments 

% Device description.
% Figure 1.
\begin{figure*}[!t]
\centering
\includegraphics[width=0.55\textwidth]{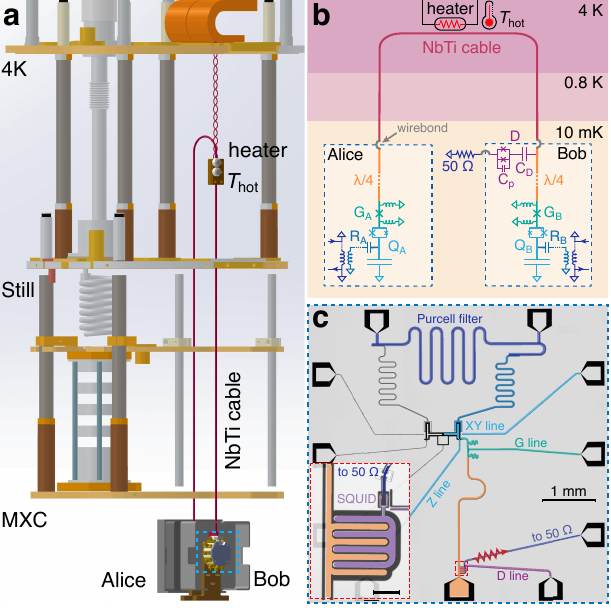}
\caption{
    {\bf Experimental setup.}
    {\bf a}, {\bf b}, Physical model and schematic of the experimental setup, where the quantum chips Alice and Bob are housed within the MXC stage at 10~mK, and the 1-m NbTi cable is hung up close to the 4~K stage of a DR. A heater and a thermometer are attached to the center of the cable.
    The communication channel is connected to a 50~$\Omega$ load at 10~mK via a tunable coupler $D$ on Bob for radiative cooling.
    {\bf c}, Photograph of Bob, showing the qubit, gmon and $D$ coupler, together with the control and readout wiring circuitries.
    Inset: zoomed in micrograph of the $D$ coupler, where the scale bar represents 50~$\mu$m.
}
\label{fig1}
\end{figure*}

\section*{Results}
% \subsection{Experimental setup}

The microwave quantum network consists of two superconducting chips Alice and Bob, connected with a 1-meter-long niobium-titanium (NbTi) cable, see the experimental setup in \rfigs{fig1}{a} and {\bf b}.
The quantum chips are situated within the mixing chamber (MXC) stage of a DR with a base temperature of $T_{\rm{cold}}\approx 10$~mK.
The cable is hung up above the still stage, close to the 4~K stage of the DR. 
A heater and a thermometer are attached to the center of the cable, allowing us to heat up the cable center and monitor its temperature change, which is denoted as $\Thot$.
The superconducting chips are fabricated with niobium, except that the Josephson junctions are made by aluminum and its oxide.
Two transmon qubits~\cite{Koch2007} are fabricated on each chip, one of which is not used; the other, denoted as $Q_n$ ($n=A$, $B$ for Alice and Bob respectively) maintains a galvanic connection to the cable through a gmon tunable coupler~\cite{Chen2014}, denoted as $G_n$ ($n=A$, $B$), and a 2.8-mm coplanar waveguide (CPW) line, see \rfig{fig1}{b}.
The CPW line serves as a quarter-wavelength ($\lambda/4$) impedance transformer centered at around 7~GHz that minimizes the cable standing-mode current at the wire bond joint~\cite{Niu2023}.
We employ a Purcell filter~\cite{Jeffrey2014} on each chip to enhance the decay rate of the readout resonator $R_n$ ($n=A$, $B$) whose resonant frequency is $\omega_{rr}^n/2\pi\sim 6$~GHz, while preserving the coherence of the qubit whose operating frequency is $\omega_n/2\pi \sim 7.5$~GHz, outside the passband of the filter.
Near the bonding pad on the Bob chip, the cable is connected to a 50~$\Omega$ microwave load anchored to $T_{\rm{cold}}$ via a tunable coupler $D$~\cite{Chang2020} for radiative cooling, here ``D'' refers to dissipation.
As shown in \rfig{fig1}{c} inset, the $D$ coupler consists of an interdigitated capacitor $C_D\approx 38.5$~fF in series with a superconducting quantum interference device (SQUID). The SQUID can be viewed as a tunable inductor $L_D = L_{D,0}/\cos(\pi \Phi_D/\Phi_0)$ shunted by the junction parasitic capacitance $C_p\approx 17$~fF, here $\Phi_D$ is the flux threading through the SQUID loop, $\Phi_0=2.067\times 10^{-15}$ is a flux quantum, $L_{D,0}\approx 3.8$~nH is the SQUID inductance at zero flux.
By tuning the SQUID flux, the impedance of the $D$ coupler can be adjusted from a maximum value (the ``off'' point) to a minimum value (the ``on'' point).
Further details of the device and the experimental setup are available in Supplementary Information.

% \subsection{Tunable channel dissipation}

% Figure 2.
\begin{figure*}[t]
\centering
\includegraphics[width=0.8\textwidth]{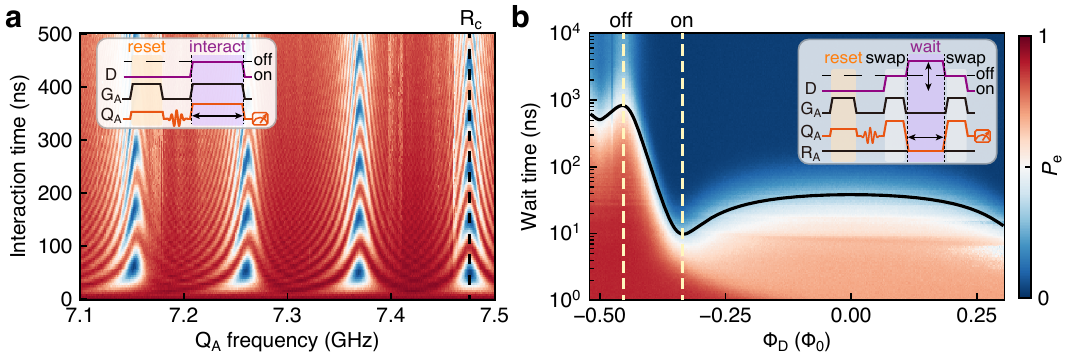}
\caption{
    {\bf Tunable channel dissipation.}
    {\bf a}, Vacuum Rabi oscillations between $Q_A$ and the standing modes.
    {\bf b}, Photon lifetime measurement for the 69-th mode at 7.48~GHz in {\bf a}, denoted as $R_c$.
    The black solid line marks the fitted photon lifetime at various $\Phi_D$.
    At $-0.46\Phi_0$, the $D$ coupler is turned ``off'', yielding a maximum photon lifetime of 820~ns.
    At $-0.32\Phi_0$, the $D$ coupler is tuned ``on'', resulting in a minimum photon lifetime of 9.6~ns.
    Insets are control pulse sequences.
}
\label{fig2}
\end{figure*}
 
Without applying electric current to the cable heater,  $\Thot=0.83$~K, in equilibrium with the still stage. Thermal noise is still weak at this moderate temperature, allowing for straightforward characterization of the communication channel with a qubit.
When coupler $G_A$ and $G_B$ are tuned off, the channel is essentially shorted to ground on both ends, supporting a series of standing bosonic modes with a free spectral range (FSR) of $\omega_{\rm{FSR}}/2\pi \approx 108$~MHz determined by its length.
The whole system can be described with the Hamiltonian:
\begin{align}
    H/\hbar = &\sum_{n=A,B}(
    \omega_n \sigma_n^{\dagger } \sigma_n 
    +\frac{\eta}{2} \sigma_n^{\dagger } \sigma_n^{\dagger } \sigma_n \sigma_n
    )
    + \sum_{m} \omega_m a_m^\dag a_m \nonumber \\ &
    + \sum_{n=A,B}\sum_{m}g_n^m(\sigma_n^\dag a + \sigma_n a^\dag),
\end{align}
here $\hbar$ is the reduced Planck constant, $\sigma_n$ and $a_m$ are the annihilation operators for qubit $Q_n$ and the $m$-th mode respectively, $\omega_m/2\pi$ is the mode frequency, $g_n^m$ is the qubit-mode coupling strength mediated by $G_n$.
The qubits are approximated as Duffing oscillators with anharmonicity $\eta$.
We prepare $Q_A$ to its first excited state $|1\rangle$ with a $\pi$ pulse following the qubit reset (see Supplementary Information for details), then turn on $G_A$ to $g_A^m/2\pi \approx 5$~MHz, and vary the qubit frequency to interact with the standing modes, during which the $D$ coupler is turned to the ``off'' point.
The qubit excitation probability $P_e=1-P_0$ is shown in \rfig{fig2}{a} where vacuum Rabi oscillations are clearly observed when $Q_A$'s frequency matches each mode.
Here $P_0$ is the qubit ground state population,  $P_e$ is the population summation of multiple higher qubit energy levels, which can be simultaneously excited with strong thermal noise and can be easily distinguished from the ground state in the quadrature space of dispersive readout, see Supplementary Information for details.

Now we focus on the $m=69$-th mode at $\omega_m/2\pi=7.48$~GHz, denoted as $R_c$, and explore its tunable dissipation controlled by the $D$ coupler.
In \rfig{fig2}{b}, we measure the photon lifetime for this mode under various $\Phi_D$ by swapping a photon from $Q_A$ to $R_c$ and waiting for certain duration, during which $Q_A$ is reset to its ground state through its readout resonator $R_A$ (see Supplementary Information for details), then retrieving back the photon for detection.
When $\Phi_D$ is biased near $-0.46\Phi_0$, the SQUID forms plasma resonance with its parasitic capacitance $C_p$, where the coupler presents a large series impedance and effectively isolates the 50~$\Omega$ load from the channel.
We measure a maximum photon lifetime of 820~ns at this $D$ coupler ``off'' point, which should be limited by the intrinsic decay rate of the mode.
Conversely, when $\Phi_D$ is biased near $-0.32\Phi_0$, the SQUID forms a series resonance with $C_D$, where the coupler presents minimal impedance, and the channel is essentially directly shorted to the cold load.
We observe a minimum photon lifetime of 9.6~ns at this $D$ coupler ``on'' point, with a two-order-of-magnitude on/off ratio.
The lifetime of other modes is also measured, showing similar performance, see Supplementary Information.

% \subsection{Radiative cooling and rethermalization}
% Figure 3.
\begin{figure*}[t]
\centering
\includegraphics[width=0.85\textwidth]{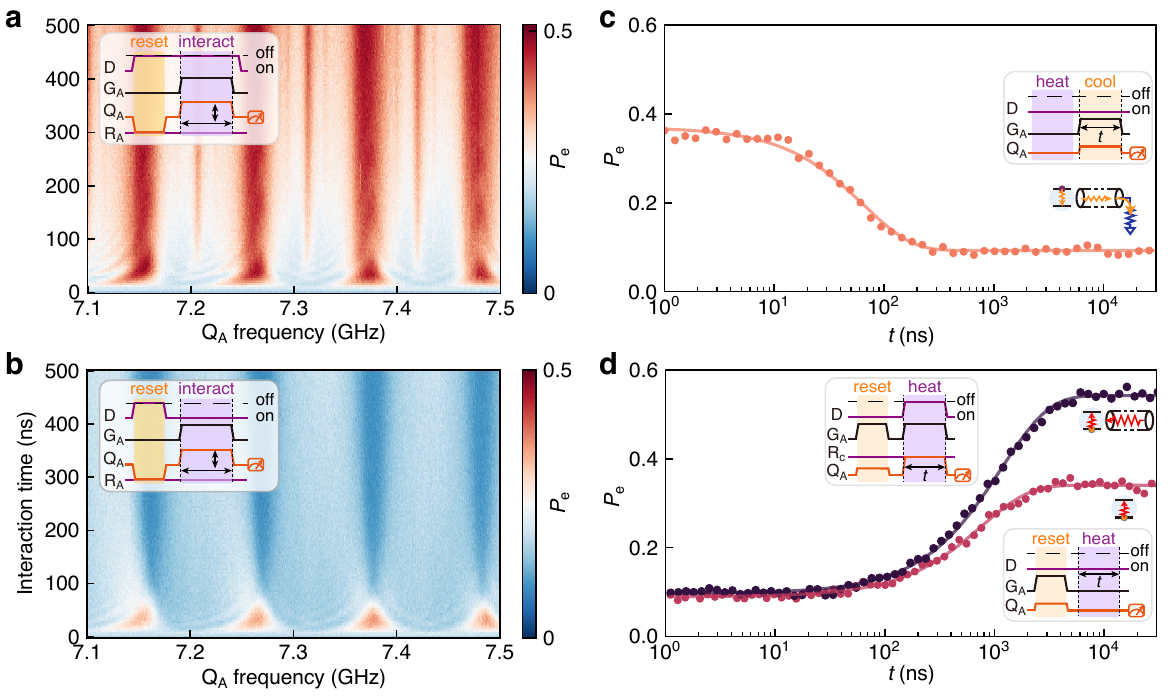}
\caption{
    {\bf Radiative cooling and rethermalization.}
    {\bf a}, Detect thermal photons in the channel with $Q_A$ at $\Thot=4$~K, when the $D$ coupler is turned ``off''.
    {\bf b}, Thermal photons are quickly reduced in 100~ns over a large frequency range after suddenly turning the $D$ coupler ``on''.
    {\bf c}, Cooling $Q_A$ by interacting with the radiatively cooled mode $R_c$. The qubit reaches a steady state with excitation $P_e = 0.095$ when at equilibrium with the mode.
    {\bf d}, Rethermalization of $Q_A$ alone, as well as $Q_A$ interacting with mode $R_c$ when $D$ coupler is suddenly turned ``off'', with a characteristic time of 731~ns and 1138~ns respectively. 
    Insets are control pulse sequences.
}
\label{fig3}
\end{figure*}

The strong dissipation at the cold load can be used to radiatively cool the channel, and reset the qubits.
To illustrate its efficiency, we heat up the cable, gradually raising $\Thot$ up to 4~K.
The thermal occupation of the $m$-th standing mode follows Bose-Einstein distribution~\cite{Jin2015}:
  $\langle N_e\rangle = 1/(e^{\hbar\omega_m/k_B T_{e}}-1)$,
here $T_{e}$ is the environmental temperature, $k_B$ is the Boltzmann constant.
Since the cable experiences varying temperatures from top to bottom, $T_{e}$ lies between $\Thot$ and $T_{\rm{cold}}$.
Superconducting qubits based microwave thermometry has been demonstrated at sub-Kelvin temperatures~\cite{Scigliuzzo2020,Wang2021,Lvov2024}, but extending these methods to liquid helium temperature is challenging.
Steady state analysis shows that $T_{e}$ is roughly the average of $\Thot$ and $T_{\rm{cold}}$, i.e., 2~K, where $\langle N_e\rangle \sim 5$ for the modes at around 7.5~GHz (see Supplementary Information for details).
At such elevated temperature, the qubit can be easily heated up by the thermal noises from the channel, as shown in \rfig{fig3}{a}.
To illustrate the process of radiative cooling, we focus on the mode $R_c$, which has a tunable coupling rate $\kappa_D$ with the 50~$\Omega$ cold load mediated by the $D$ coupler as shown in \rfig{fig2}{b}. The mode occupancy at equilibrium becomes~\cite{Albanese2020,Xu2020,Wang2021}
\begin{equation}
    \langle N_c\rangle = \frac{\kappa_i \langle N_{e}\rangle + \kappa_D \langle N_{\rm{cold}}\rangle}{\kappa_i + \kappa_D} \approx \frac{1}{1 + \kappa_D/\kappa_i}\langle N_{e}\rangle,
\end{equation}
here $\kappa_i$ is the mode's intrinsic coupling to the warm environment, $\langle N_{\rm{cold}}\rangle\approx 0$ is the thermal occupation of the cold load.
When $\kappa_D/\kappa_i \gg 1$, we have $\langle N_c\rangle \ll \langle N_{e}\rangle$ as a consequence of radiative cooling.
Assuming the maximum photon lifetime in \rfig{fig2}{b} is limited by the mode's intrinsic coupling to the environment, we estimate that $\kappa_i\approx 1/820$~ns$^{-1}$ at $\Thot=0.83$~K.
As the thermal occupation far exceeds one at $\Thot=4$~K, it becomes impractical to calibrate the mode $\kappa_i$ using a qubit.
Independent tests show that the NbTi cable maintains a stable quality factor above $10^5$ up to several Kelvins (see Supplementary Information for details), suggesting the change of $\kappa_i$ is insignificant within the temperature range of interest, leaving $\kappa_D$ as the predominant factor.
The cooling effect can be clearly manifested in \rfigs{fig3}{a} and {\bf b}, as we dynamically tune $D$ from ``off'' to ``on'', yielding a maximum $\kappa_D\approx 1/9.6$~ns$^{-1}$ assuming the maximum coupling rate is hardly affected by thermal noise.
We estimate that the number of thermal photons in the channel is reduced by a factor of $1+\kappa_D/\kappa_i\approx 86$, reaching a noise floor of $\langle N_c\rangle \sim 0.06$ photons after radiative cooling (see Supplementary Information for details).
The $D$ coupler can also reset the qubit through the channel, as shown in \rfig{fig3}{c}.
Without reset, the qubit has a steady excitation of $P_e=0.34$ at $\Thot=4$~K.
Once we turn on the $G_A$ coupler, thermal occupation in $Q_A$ decreases, reaching a steady excitation of $P_e=0.095$ when at equilibrium with the radiatively cooled mode $R_c$.

As long as the $D$ coupler is kept ``on'', the channel remains quiet but lossy.
To transfer microwave quantum states through the channel, the $D$ coupler needs to be temporarily turned ``off'' to avoid loss of coherent photons.
Then the channel as well as the qubits rethermalize immediately, the rate of which is vital for the transfer process to complete.
In \rfig{fig3}{d} (red data), we display the ground-state thermalization of $Q_A$ alone, with $G_A$ turned off after the qubit reset.
$Q_A$ finally thermalizes to a steady state with $P_e=0.34$ at equilibrium, with a characteristic time of $\tau = 760$~ns (see Supplementary Information for details).
To monitor the channel rethermalization, we use $Q_A$ to interact with the mode $R_c$ by tuning their coupling strength to $g_A^c/2\pi\approx 5$~MHz, when coupler $D$ is suddenly turned off.
As shown in \rfig{fig3}{d} (dark red data), the qubit reaches a steady state with $P_e=0.556$ at equilibrium, with a characteristic time of $\tau = 1.14$~$\mu$s.
This steady state excitation is significantly higher compared to that of the qubit alone, because the mode is hotter than the qubit.
Numerical simulation shows that this steady state excitation is related to the number of thermal photons in the mode, from which we estimate $\langle N_e\rangle\sim 5$, corresponding to $T_e\sim 2$~K, roughly the average of $\Thot$ and $T_{\rm{cold}}$.
This is perhaps not fortuitous, given the inner conductor of the cable is not thermalized to intermediate stages in the DR. We also benchmark the radiative cooling and rethermalization processes at lower channel temperatures, see Supplementary Information.

% \subsection{Quantum state transfer and remote entanglement}

% Figure 4.
\begin{figure*}[t]
\centering
\includegraphics[width=0.98\textwidth]{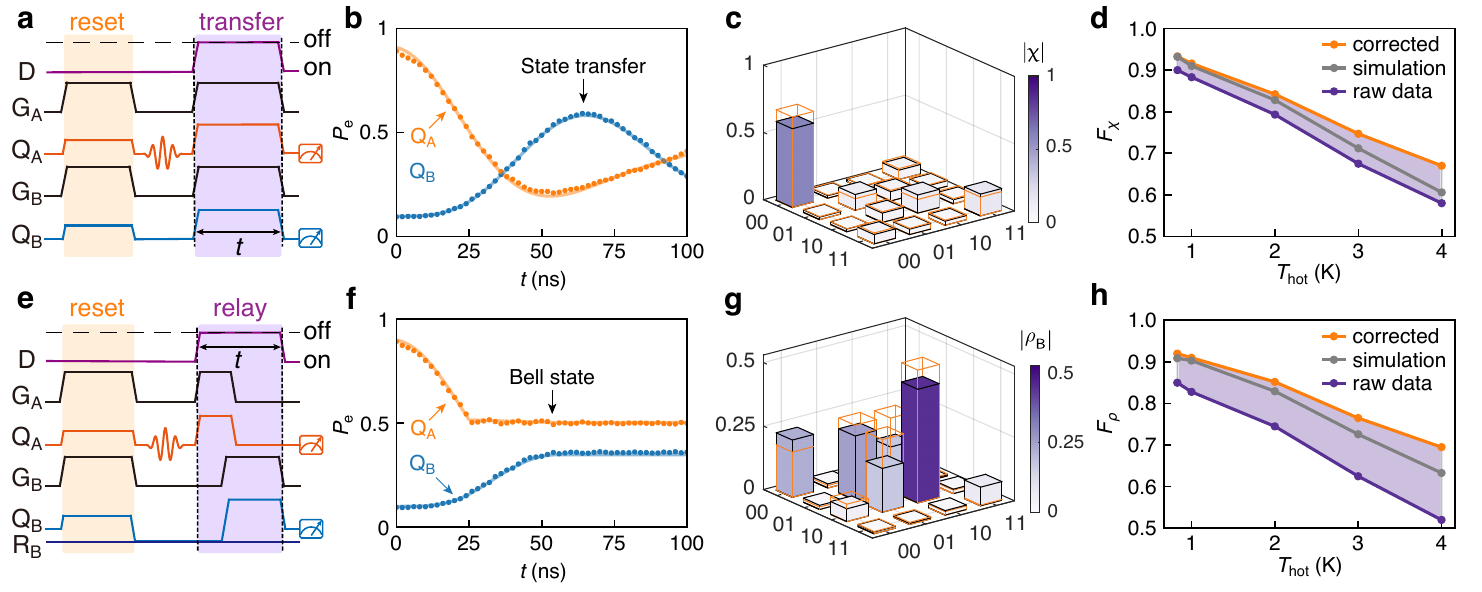}
\caption{
    {\bf Quantum state transfer and remote entanglement.}
    {\bf a}, {\bf b}, Protocol and data for transferring a photon from $Q_A$ to $Q_B$ through $R_c$.
    {\bf c}, Process matrix $\chi$ for the quantum state transfer in {\bf b} at $t = 65$~ns, with fidelity $\mathcal{F}_{\chi} = 58.5\pm 0.5\%$ without (purple bars) and $67.2\pm 0.7\%$ with SPAM error correction (orange frames).
    {\bf d}, $\mathcal{F}_\chi$ at different $\Thot$ spanning from 0.83~K to 4~K, where purple and orange dots are data without and with SPAM error correction, respectively. Gray dots are numerical simulation results.
    {\bf e}, {\bf f}, Protocol and data for transferring half a photon from $Q_A$ to $Q_B$.
    {\bf g}, Density matrix $\rho_B$ for the Bell state generated at $t = 53$~ns, with fidelity $\mathcal{F}_\rho = 52.3\pm 0.5\%$ without (purple bars) and $69.6\pm 0.7\%$ with SPAM error correction (orange frames).
    {\bf h}, $\mathcal{F}_\rho$ at different $\Thot$, similar to {\bf d}.
}
\label{fig4}
\end{figure*}

After reducing the thermal noise to $\sim 0.06$ photons, we can transfer microwave quantum states through the channel.
In Figs.~\ref{fig4}{\bf a} and {\bf b}, we demonstrate the transfer of a single photon from $Q_A$ to $Q_B$ through $R_c$, by simultaneously turning on $G_A$ and $G_B$ to an equal coupling strength of $g_A^c = g_B^c \approx 2\pi\times 5$~MHz for a duration of $t$~\cite{Wang2012,Zhong2021}, meanwhile turning $D$ coupler ``off''.
The standing mode near 7~GHz shows slightly better coherence as it matches the $\lambda/4$ impedance transformer (see Supplementary Information), but we choose $R_c$ for communication because its frequency is closer to the qubit operating frequency.
The state transfer process is completed at $t = 65$~ns, an order of magnitude smaller than the rethermalization characteristic time in Fig.~\ref{fig3}{\bf d}.
Figure~\ref{fig4}{\bf c} illustrates the process matrix $\chi$ for the transfer process, measured by quantum process tomography~\cite{Zhong2021}, yielding a fidelity $\mathcal{F}_{\chi}=\textrm{Tr}(\chi\cdot \mathcal{I})=58.5 \pm 0.5\%$, here $\mathcal{I}$ is the identity operation.
All uncertainties reported here represent the standard deviation of repeated measurements.
We note that the qubit readout is degraded as the channel becomes hot, and the fidelities calculated from the raw data are underestimated.
If we correct the data for the state preparation and measurement (SPAM) errors, we obtain $\mathcal{F}_\chi = 67.2\pm 0.7\%$ (see Supplementary Information).
For comprehensive investigation, The process fidelity $\mathcal{F}_{\chi}$ at various $\Thot$ spanning from 0.83~K to 4~K are also measured, as shown in Fig.~\ref{fig4}{\bf d}.
Remote entanglement is generated by transferring half a photon from $Q_A$ to the mode first, then relaying it to $Q_B$, see Figs.~\ref{fig4}{\bf e} and {\bf f}, resulting in a Bell state $|\psi_B\rangle=(|01\rangle+|10\rangle)/\sqrt{2}$ between $Q_A$ and $Q_B$ ideally.
Figure~\ref{fig4}{\bf g} shows the density matrix $\rho_B$ of the Bell state, with a fidelity of $\mathcal{F}_\rho = \langle \psi_B|\rho_B|\psi_B\rangle =52.3\pm 0.5\%$ for raw data, and $69.6\pm 0.7\%$ if corrected for SPAM errors.
Numerical simulation yields a Bell state fidelity of $63.3\%$, lying between the raw and corrected data fidelities.
The Bell state fidelities at various $\Thot$ are also measured, as shown in Fig.~\ref{fig4}{\bf h}.
The corrected data fidelities agree with simulation well at low temperatures, as expected; but when the channel becomes warmer, higher qubit energy levels get involved that the SPAM error correction becomes less accurate.

%\section{Summary and outlook}
\section*{Conclusion}
Our demonstration of thermal-noise-resilient microwave quantum network redefines the operational constraints for superconducting quantum systems.
This thermal compatibility enables quantum state transfer through conventional cryogenic infrastructure while maintaining photon-mediated entanglement fidelity above classical thresholds, overcoming a key obstacle in modular quantum architecture design~\cite{Habraken2012,Xiang2017,Vermersch2017,Wehner2018}.
By operating the communication channel up to 4~K--two orders of magnitude warmer than current quantum coherent links--we establish a bridge between millikelvin quantum processors and liquid-helium temperature regimes where hybrid quantum systems~\cite{Xiang2013,clerk2020,Huang2024a} and microwave-optical transducers~\cite{Mirhosseini2020,Forsch2020,Tu2022,Kumar2023,Sahu2023,Jiang2023,Meesala2024a} may reside. 
Building on recent advancements in hot silicon qubits~\cite{petit2020,yang2020,camenzind2022,huang2024} and their integration with microwave quantum circuits~\cite{dijkema2025}, our work extends the feasibility of hybrid quantum networks to semiconductor-based platforms.

The achieved fidelities, currently limited by residual broadband thermal noise and qubit coherence degradation, can be systematically improved through these parallel advances: (1) integration of cryogenic bandpass filters~\cite{Habraken2012}, niobium junctions~\cite{Lisenfeld2007,Anferov2024a,Anferov2024} and quasiparticle traps~\cite{Sun2012,Wang2014} to suppress off-resonant noise, (2) adoption of higher quality superconducting cables~\cite{Niu2023,Qiu2024,Mollenhauer2024}, (3) implementation of error-corrected transfer protocols~\cite{Burkhart2021,Xiang2017,Stas2022} and linear transceivers for multi-photon transmission~\cite{Grebel2024}, and (4) application of new cooling techniques, such as sideband cooling~\cite{Valenzuela2006,Gely2019,Rodrigues2021} and quantum thermal machines~\cite{Aamir2025}. %Recent progress in aluminum-based quantum coaxial lines~\cite{} suggests equivalent NbTi implementations could simultaneously reduce photon loss and thermalization rates by over an order of magnitude.

Looking beyond device engineering, our results establish thermal microwave networks as a testbed for quantum thermodynamics. The observed non-Markovian dynamics during channel rethermalization invite theoretical exploration of quantum thermodynamics~\cite{pekola2015,rossnagel2016,ronzani2018,pekola2021,lu2022a,pekola2022} in engineered cryogenic environments. Practical applications will benefit from synergies with photonic quantum technologies--our approach relaxes the stringent cooling requirements for microwave-optical transduction~\cite{Kimble2008,Tu2022}, potentially enabling room-temperature optical links to interface with superconducting processors through thermally buffered microwave channels~\cite{Aspelmeyer2014,Jiang2023}.

\clearpage

\bibliographystyle{naturemag}
\bibliography{ref}

%\clearpage
%%%%% END MATTER

\section*{Acknowledgements}
We thank Jian-Wei Pan and Luming Duan for insightful discussions, and thank Hai Jin for providing thermometers used in this experiment.
This work was supported by the Science, Technology and Innovation Commission of Shenzhen Municipality (KQTD20210811090049034), the National Natural Science Foundation of China (12174178, 12374474), the Innovation Program for Quantum Science and Technology (2021ZD0301703), and Guangdong Basic and Applied Basic Research Foundation (2024A1515011714, 2022A1515110615).

\section*{Author contributions}
Y.Z. conceived the experiment.
J.Q. designed and fabricated the devices assisted by L.Z.
J.Q. and Z.Z. performed the measurements and analyzed the data guided by J.N.
Y.Z. and D.Y. supervised the project.
All authors contributed to discussions and production of the manuscript.

\section*{Competing interests}
The authors declare no competing interests.

\section*{Data availability}
The data that support the plots within this paper and other findings of this study are available from the corresponding author upon reasonable request.

\end{document}

%\linenumbers
\title{Supplementary Information for ``A thermal-noise-resilient microwave quantum network traversing 4~K'}

\maketitle
\tableofcontents
\setcounter{equation}{0}
\setcounter{figure}{0}
\setcounter{table}{0}
\setcounter{page}{1}

\renewcommand{\figurename}{Fig.}
\newcommand{\rfig}[2]{Fig.~\ref{#1}{\bf #2}}
\newcommand{\rfigs}[2]{Figs.~\ref{#1}{\bf #2}}
\newcommand{\req}[1]{Eq.~(\ref{#1})}
\newcommand{\reqs}[1]{Eqs.~(\ref{#1})}
\renewcommand{\theequation}{S\arabic{equation}}
\renewcommand{\thefigure}{S\arabic{figure}}
\renewcommand{\thetable}{S\arabic{table}}
\newcommand\Thot{T_\mathrm{hot}}
\newcommand\Tcold{T_\mathrm{cold}}
\newcommand\Tr{T_\mathrm{1r}}
\newcommand\Th{T_\mathrm{h}}
\newcommand\XXX{\textcolor{red}{XXX}}
\newcommand\red{\textcolor{red}}

\newpage

\section{Experimental setup}\label{sec:setup}

\begin{figure}[H]
\centering
\includegraphics[width=0.7\textwidth]{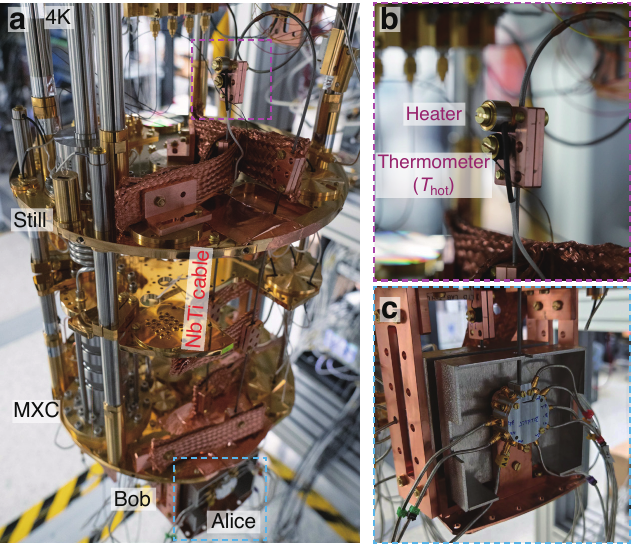}
\caption{
    \label{fig_setup}
    {\bf Experimental setup.} 
    {\bf a}, Photograph of the setup inside a dilution refrigerator. 
    {\bf b}, Zoomed in photograph of the heater and thermometer anchored near the top of the cable.
    {\bf c}, Zoomed in photograph of Alice inside a magnetic shield. Bob is positioned behind it.
}
\label{setup}
\end{figure}

The experimental setup is housed within a BlueFors LD400 dilution refrigerator (DR), as shown in \rfig{fig_setup}{\bf a}.
We use a 1-meter-long niobium-titanium (NbTi) superconducting coaxial cable from Keycom Corp. to connect two superconducting processors, Alice and Bob. The cable is bent into U-shape and vertically mounted in the DR. With an outer diameter of 2.2~mm, the cable possesses sufficient mechanical rigidity to stand upright without additional mechanical support. The cable traverses through the mixing chamber (MXC) stage, the cold plate (CP), and the still stage, close to the 4~K stage of the DR.
Near its apex, a heater and a thermometer are anchored to the cable using a copper clamp, as shown in \rfig{fig_setup}{b}. 
The heater, powered by a DC source at room temperature, enables controlled heating of the cable center, while the thermometer monitors the local temperature, $\Thot$, in real time. In the absence of heating, $\Thot$ stabilizes at approximately 0.83 K, in equilibrium with the still plate. By applying heating power up to 52.5 microwatts, $\Thot$ can be continuously adjusted from 0.83~K to 4.0~K, with fluctuations less than 0.1~K after stabilization. 
The low thermal conductivity of NbTi at cryogenic temperatures ensures that the temperatures of the still plate and other parts in the DR remain largely unaffected by the heating.

\begin{figure}[H]
  \centering
  \includegraphics[width=0.9\textwidth]{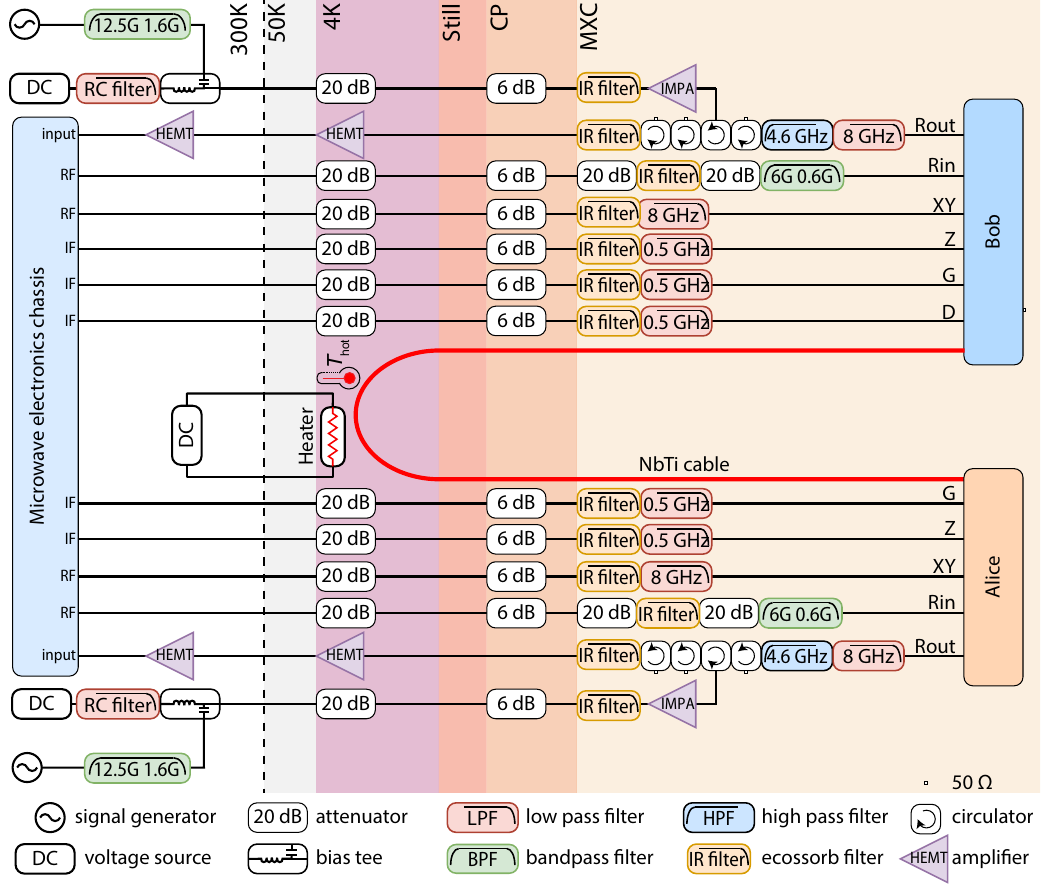}
  \vspace{-10pt}
  \caption{\label{fig_wiring}
  {\bf Room-temperature electronics and cryogenic wiring diagram.}}
\vspace{-10pt}
\end{figure}

The quantum processors are enclosed within magnetic shields and anchored to the MXC stage with copper brackets, as shown in \rfig{fig_setup}{c}.
Positioned back-to-back, each processor is connected to multiple microwave coaxial cables that facilitate the transmission of control and measurement signals.
A custom-designed microwave electronics chassis~\cite{zhang2024} is employed for qubit control and readout.
An overview of the electronics configuration and cryogenic wiring is shown in \rfig{fig_wiring}{}.

\section{Device information}

The design, measured parameters, and typical performance of each device component are summarized in Table \ref{tab_parameters}. More details are provided in the subsections below.

\begin{table}[H]
\centering
\begin{tabular}{l c c c}
\hline \hline
$D$ coupler parameters                                                    & $D$        &        & \\
\hline
coupler series capacitance, $C_D$ (fF)                                    & 38.5       &        & \\
coupler parasitic capacitance, $C_p$ (fF)                                 & 17         &        & \\
coupler SQUID inductance, $L_{D,0}$ (nH)                                  & 3.8        &        & \\
coupler parasitic resistance, $R_1$ ($\Omega$)                            & 2300       &        & \\
minimum dissipation rate, $\kappa_D^{\mathrm{min}}$  ($\mathrm{ms}^{-1}$) & 1/1.7      &        & \\
maximum dissipation rate, $\kappa_D^{\mathrm{max}}$  ($\mathrm{ns}^{-1}$) & 1/9.6      &        & \\
\hline \hline
qubit parameters                                                          & $Q_A$      & $Q_B$  & \\
\hline
min freq., $\omega_{q}^{\rm{min}}/2\pi$ (GHz)                             & 4.604      & 4.634  & \\
max freq., $\omega_{q}^{\rm{max}}/2\pi$ (GHz)                             & 7.502      & 7.723  & \\
idling freq., $\omega_{q}/2\pi$ (GHz)                                     & 7.429      & 7.538  & \\
anharmonicity, $\eta/2\pi$ (MHz)                                          & $-204$     & $-204$ & \\
readout resonator freq., $\omega_{rr}/2\pi$ (GHz)                         & 5.963      & 5.967  & \\
readout time (ns)                                                         & 190        & 190    & \\
\hline \hline
gmon coupler parameters                                                   & $G_A$      & $G_B$  & \\
\hline
gmon junction inductance (nH)                                             & 0.6        & 0.6    & \\
gmon grounding inductance (nH)                                            & 0.2        & 0.2    & \\

\hline \hline
channel parameters                                                        & NbTi cable & CPW    & \\
\hline
inductance per unit length, $\mathcal{L}_j$ (nH/m)                        & 240.5      & 402    & \\
capacitance per unit length, $\mathcal{C}_j$ (pF/m)                       & 96.2       & 173    & \\
length (m)                                                                & 1          & 0.0028 & \\

\hline
\hline
\end{tabular}
\caption{\label{tab_parameters} {\bf Device parameters and typical performance.}}
\end{table}

The superconducting quantum chips for Alice and Bob share almost identical design, except that Alice does not have a $D$ coupler. Both chips are fabricated from the same batch of sapphire wafer, on which a 100~nm-thick niobium thin film is deposited via magnetron sputtering at room temperature.
Most of the circuit elements are patterned on this niobium layer using lithography, followed by wet etching with an aqueous solution of hydrofluoric and nitric acids.
The Josephson junctions are fabricated using the Dolan bridge method~\cite{dolan1977}, with aluminum thin films deposited through electron-beam evaporation.
After fabricating the Josephson junctions, an additional aluminum layer is deposited using lithography to form galvanic connections between the junctions and the base niobium film~\cite{dunsworth2017}.
To suppress slotline modes, crossovers with $\text{SiO}_2$ scaffolds are used to connect the ground planes across the coplanar waveguides (CPWs) on the chip~\cite{chen2014a,dunsworth2018}.

\subsection{\texorpdfstring{$D$}{D} coupler}

Fast and tunable channel dissipation is crucial for our experiments involving thermal microwave photons. In our setup, this capability is realized using the $D$ coupler, here the name ``D'' refers to dissipation.
In this subsection, we provide a comprehensive description of the $D$ coupler's design and performance.

\begin{figure}[H]
\centering
\includegraphics[width=0.8\textwidth]{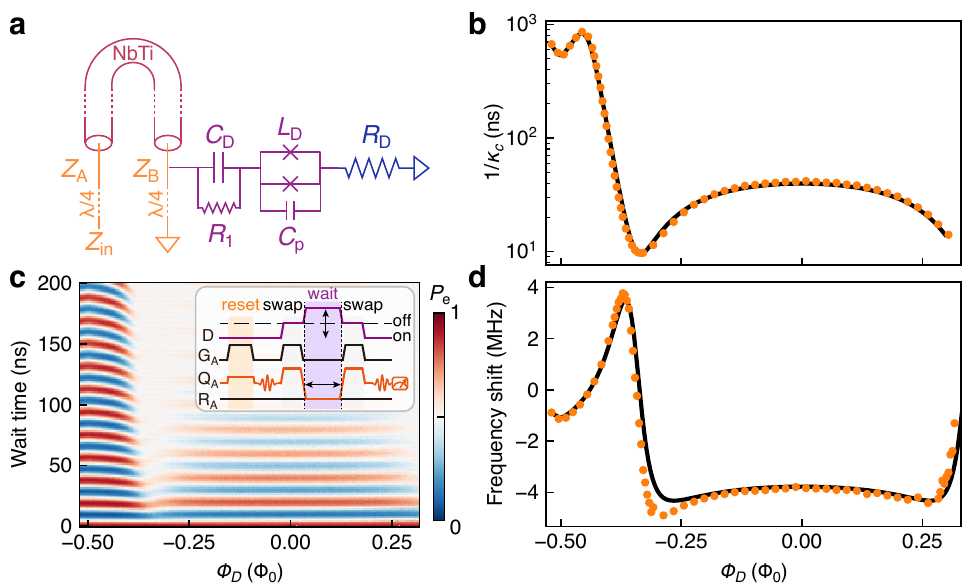}
\caption{
\label{fig_D_circuit}
    {\bf Characterization of the $D$ coupler.}
    {\bf a}, Circuit model of the $D$ coupler. The NbTi cable is connected to the $D$ coupler circuit and shorted to ground. The $D$ coupler consists of an interdigital capacitor $C_D$, a SQUID with an effective inductance $L_D$, and a parasitic capacitance $C_p$. The circuit is terminated by a $50~\Omega$ load $R_D$, anchored to 10~mK, which provides dissipation. $R_1$ is parasitic resistance to account for the finite on/off ratio of the $D$ coupler.
    {\bf b}, Mode dissipation rate $\kappa_c$ versus $\Phi_D$. The orange dots represent the characteristic time of exponential fits to Fig.~2{\bf b} of the main text. The solid line is given by calculation from the circuit model. 
    {\bf c}, Ramsey experiment traces of the mode $R_s$ at various flux bias for the $D$ coupler. Inset shows the control pulse sequence.
    {\bf d}, Frequency shifts of $R_c$ by $\Phi_D$. The orange dots are the oscillation frequencies in {\bf a} obtained by fitting. The solid line is given by calculation from the circuit model. The frequency shift is the shift compared to the $D$ coupler off point.
}
\end{figure}

As described in the main text, the $D$ coupler comprises a DC SQUID in series with an interdigitated capacitor $C_D$.
This circuit functions as a tunable bandpass filter, connecting the communication channel to a $R_D=50~\Omega$ cold load.
The SQUID serves as a tunable inductance $L_D$, which can determine the passband frequency $\omega_D \approx 1/\sqrt{L_D C_D}$.
At zero flux bias, $L_D = 3.8$~nH and the maximum $\omega_D/2\pi \approx 13$~GHz, encompassing the operating frequencies of the qubits.
Note that the junction area also introduces a parasitic capacitance $C_p\approx 17~\mathrm{fF}$. When the SQUID forms plasma resonance with $C_p$, the $D$ coupler exhibits high impedance, effectively decoupling the channel from the cold load.
Additionally, we account for a parasitic resistance $R_1$ due to the finite on/off ratio of the $D$ coupler, as well as the $\lambda/4$ transformer that defines the cable boundary. The full circuit model is illustrated in \rfig{fig_D_circuit}{a}.

We calculate the standing mode frequency and its quality factor using the circuit model by analyzing the input impedance from either side~\cite{pozar2012,Chang2020}.  For example, on the Alice side, the input impedance is determined by the CPW terminated with the load $Z_A$, which is defined by the NbTi cable and the circuitry on the Bob side:
\begin{align}
Z_\mathrm{in} &= Z_t \frac{Z_A + iZ_t\tan(\beta_t l_t)}{Z_t + iZ_A\tan(\beta_t l_t)},\\
Z_A &= Z_\mathrm{cb} \frac{Z_B + iZ_\mathrm{cb}\tan(\beta_\mathrm{cb} l_\mathrm{cb})}{Z_\mathrm{cb} + iZ_B\tan(\beta_\mathrm{cb} l_\mathrm{cb})},\\
Z_B &= \left(\frac{1}{Z_D} + \frac{1}{iZ_t\tan{\beta_t l_t}} \right)^{-1},\\
Z_D &= \left(i\omega C_D + \frac{1}{R_1} \right)^{-1}
		+ \left(\frac{1}{i\omega L_D}+ i\omega C_p \right)^{-1}
		+ R_D,
\end{align}
here $Z_j,~\beta_j,~l_j$ ($j=t, \mathrm{cb}$) are the characteristic impedance, wave vector and length of the $\lambda/4$ CPW line and the NbTi cable, respectively. The values of $Z_j, \beta_j$ are determined by the inductance per unit length $\mathcal{L}_j$, and capacitance per unit length $\mathcal{C}_j$ through
\begin{align}
Z_j &= \sqrt{\mathcal{L}_j / \mathcal{C}_j},\\
\beta_j &= \omega\sqrt{\mathcal{L}_j \mathcal{C}_j}.
\end{align}
The specific values of $\mathcal{L}_j$ and $\mathcal{C}_j$ for the NbTi cable and the CPW line are summarized in Table~\ref{tab_parameters}.

By calculating $Z_\mathrm{in}(\omega)$ as a function of the angular frequency $\omega$, we identify a series of zeros in $\mathrm{Im}[Z_\mathrm{in}(\omega)]$, corresponding to the resonance frequencies $\omega_m$ of the standing modes inductively coupled to the input port. 
% The roots correspond to the resonance frequencies, denoted by $\omega_m$. 
The quality factor $Q$ is given by~\cite{pozar2012}
\begin{align}
Q = \omega_m/\kappa_D = \omega_m 
\frac{1}{\mathrm{Re}[Z_\mathrm{in}(\omega_m)]}
\frac{1}{2}
\frac{\partial\mathrm{Im}[Z_\mathrm{in}(\omega)]}{\partial\omega}\Big{|}_{\omega=\omega_m}.
\end{align}

By tuning the bias flux $\Phi_D$ of the $D$ coupler, the inductance $L_D = L_{D,0}/\cos{(\pi\Phi_D/\Phi)}$ is modified, resulting in changes to both the dissipation rate $\kappa_D$ and the mode frequency $\omega_m$.
we consider the standing mode $R_c$ at 7.48~GHz here and take intrinsic dissipation $\kappa_i$ into account.
The total dissipation rate of the mode is given by $\kappa_c = \kappa_i + \kappa_D$ where $\kappa_i=1/820~\mathrm{ns}^{-1}$. 
The tunability of $\kappa_c$ is demonstrated by measuring the energy relaxation of $R_c$ in Fig.~2{\bf b} of the main text, from which we extract the photon lifetime $1/\kappa_c$ through exponential fitting. The results shown as orange dots in \rfig{fig_D_circuit}{b}. 

The frequency shifts are experimentally measured through Ramsey experiments conducted at various $\Phi_D$ bias, as shown in \rfig{fig_D_circuit}{c}. A superposition state $|0\rangle+|1\rangle$ is initialized on $Q_A$ and then swapped into $R_c$. After tunning $\Phi_D$ to a specific value and waiting for a varying period $t$, we swap the photon back to $Q_A$, apply a second $\pi/2$ pulse, and then read out the qubit state. The observed oscillation of the excited-state probability $P_e$ with $t$ reflects the mode frequency $\omega_c/2\pi$ with high accuracy. By fitting the data to a sinusoidal function with an exponential decaying amplitude, the frequency shifts are extracted and shown as orange dots in \rfig{fig_D_circuit}{d}. We notice that a distortion in the $D$ bias pulse introduces slight asymmetry along $\Phi_D=0$ in the measured frequency shifts.

To compare the experimental data with the predictions of the circuit model, we fit the model using the parameters $L_{D,0}=3.8~\mathrm{nH}$, $C_D=38.5~\mathrm{fF}$, $C_p=17~\mathrm{fF}$ and $R_1=2300~\Omega$. The calculated results, depicted as solid lines in \rfigs{fig_D_circuit}{b} and {\bf d}, demonstrate the consistency between the model and the measurements. 
According to this model, we estimate that the minimum dissipation rate introduced by the $D$ coupler at the ``off'' point is  $\kappa_D^{\mathrm{min}}=1.7$~$\mathrm{ms}^{-1}$, far more smaller than the intrinsic dissipation rate of the mode and can be safely ignored.

\subsection{Qubit readout and reset}

\begin{figure}[H]
\centering
\includegraphics[width=0.98\textwidth]{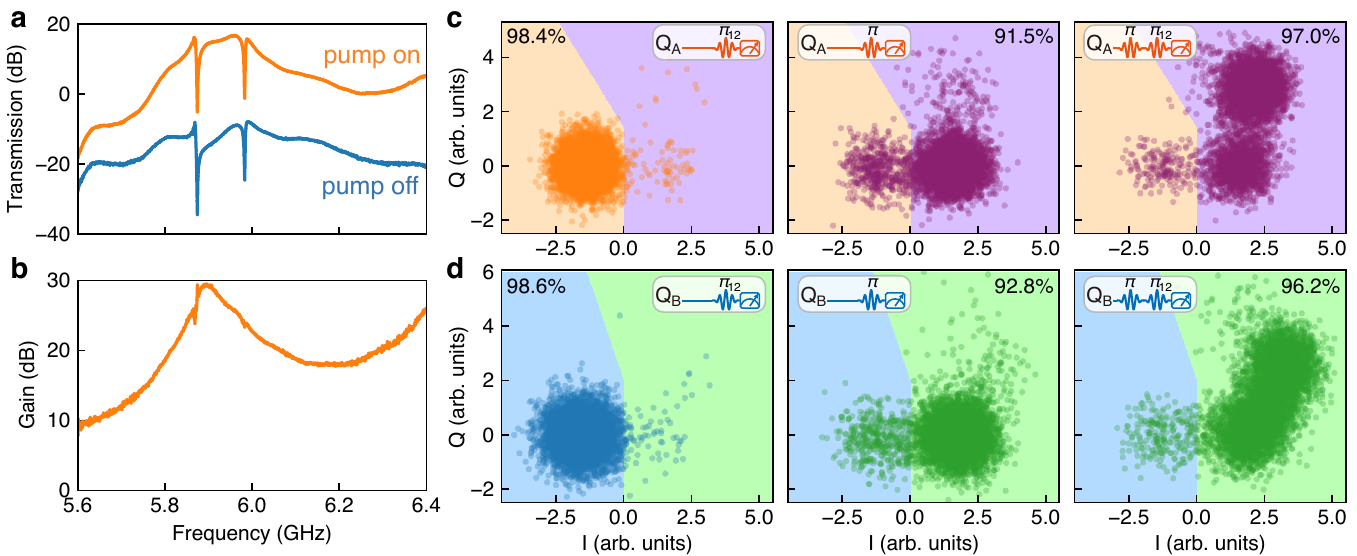}
\caption{\label{fig_readout}
    {\bf Qubit readout performance.}
    {\bf a}, Transmission spectrum of the readout feedline with/without IMPA pump. 
    The bandpass feature comes from the integrated Purcell filter, whose passband matches the readout resonators. 
    {\bf b}, Gain profile of the IMPA extracted from {\bf a}, demonstrating a gain of over 25~dB around the readout resonator frequency.
    {\bf c, d}, Single-shot readout results of $Q_A$ (c) and $Q_B$ (d) at $\Thot=0.83~\mathrm{K}$, with a readout pulse of 190~ns. The left column are IQ scatter data when the qubits are prepared in $|0\rangle$ state;
    The central column are IQ scatter data when the qubits are prepared in $|1\rangle$ state; The right column are IQ scatter data when the qubits are applied with a $\pi_{12}$ pulse to transfer the population from $|1\rangle$ to $|2\rangle$ state.
    Insets are the control sequences.
}
\end{figure}

When $\Thot$ reaches 4.0~K, thermal noises diffuse to the quantum circuits through various degrees of freedom, and the coherence time of the qubits decrease to below $1~\mathrm{\mu s}$ (see \rfig{fig_qutip_simu}{a} for details), necessitating fast qubit readout within a few hundred nanoseconds.
To achieve this, we employ a Purcell filter~\cite{satzinger2018,bienfait2019,Zhong2021} to facilitate rapid photon radiation from the readout resonator without compromising the qubit's relaxation time imposed by the Purcell effect~\cite{houck2008,Jeffrey2014,walter2017}. The filter is centered at approximately 6 GHz, with a bandwidth of more than 100~MHz.
Additionally, we employ a quantum-limited impedance-matched parametric amplifier (IMPA)~\cite{roy2015,Grebel2021} (as illustrated in the circuit diagram of \rfig{fig_wiring}{}) to amplify the readout signal.
The transmission spectrum of the readout feedline for Alice is shown in \rfig{fig_readout}{a}. With appropriate bias and pump power applied to the IMPA, the readout signal is amplified by approximately 25 dB over a bandwidth of 100 MHz, as shown in \rfig{fig_readout}{b}. Similar performance is also achieved for the IMPA of Bob. 

\begin{figure}[H]
\centering
\includegraphics[width=0.8\textwidth]{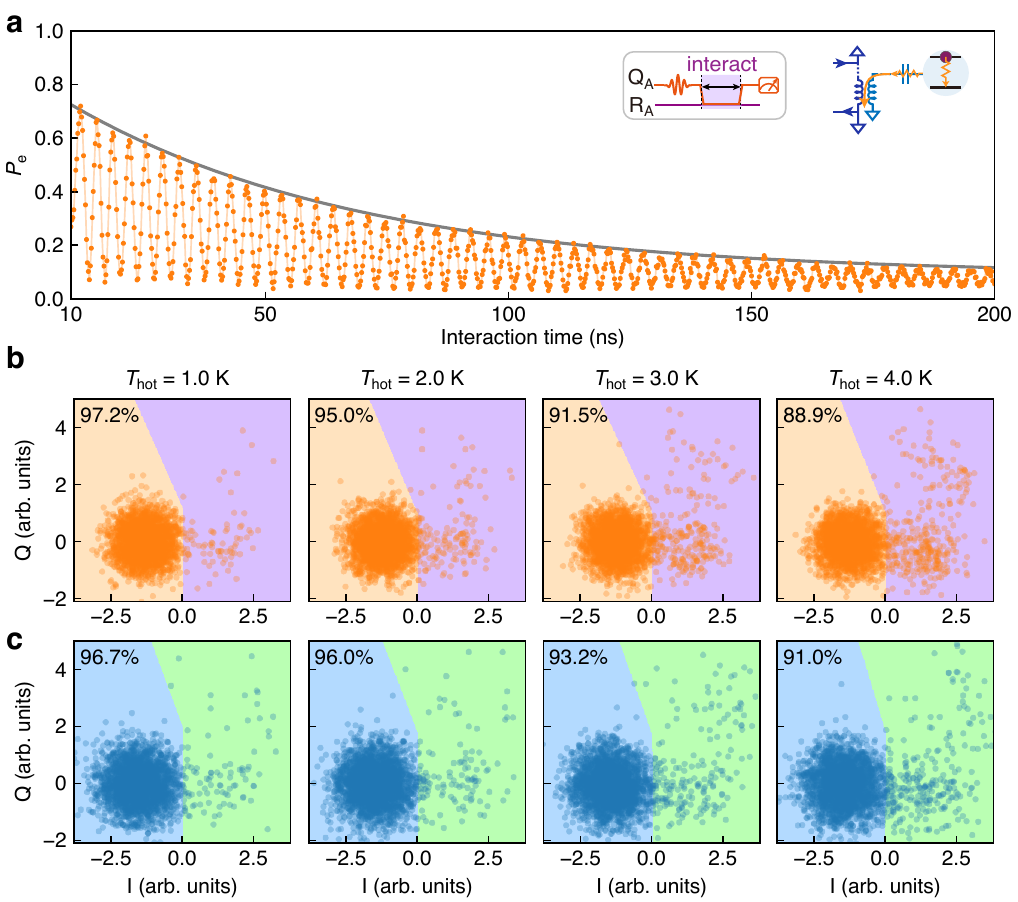}
\caption{\label{fig_readout_reset}
    {\bf Qubit reset with readout resonator.}
    {\bf a}, Interaction between $Q_A$ and the readout resonator $R_A$. The dots represent experimental data and the solid line shows the fit with a simple exponential function. Inset shows the control pulse sequence and the energy flow in this experiment.
    {\bf b, c}, Single-shot readout results of $Q_A$ ($Q_B$) after reaching equilibrium with readout resonator $R_A$ ($R_B$). Columns show data at different $\Thot$. Annotations indicate the proportion of results discriminating qubit at $|0\rangle$.
}
\end{figure}

By combining the Purcell filter with the IMPA, we achieve rapid single-shot dispersive readout for both $Q_A$ and $Q_B$ within 190~ns, maintaining a separation error below 1\%. However, the state preparation and measurement (SPAM) error remains significant, partly due to energy decay during the readout process. To further enhance the readout fidelity, we leverage the second excited state $|2\rangle$ to assist the qubit readout~\cite{martinis2002,mallet2009,chen2023}. Before applying the readout pulse, a $\pi_{12}$ pulse is applied to transfer the population from the $|1\rangle$ state to the $|2\rangle$ state.
As the $|2\rangle$ state decays to $|1\rangle$ state first, then to the $|0\rangle$ state, this results in a longer process for the qubit to decay to the $|0\rangle$ state, thus enhancing the readout fidelity.
\rfigs{fig_readout}{c} and {\bf d} show the single-shot readout results for $Q_A$ and $Q_B$ after prepared in the $|0\rangle$, $|1\rangle$ and $|2\rangle$ states, respectively.
Leveraging the $|2\rangle$ state and discriminating between the $|0\rangle$ state and all the excited states (including $|1\rangle$ and $|2\rangle$), we achieve a reduced SPAM error below 4\%.
The qubit readout performance at elevated temperatures are shown in \rfig{fig_readout_reset}{} and \rfig{fig_iqs}{}.

%\subsection{Qubit reset with readout resonator}

As demonstrated in Fig.~3 in the main text, the $D$ coupler can effectively cool/reset both the cable modes and the qubits. However, there are scenarios where it is necessary to reset the qubits independently without perturbing the cable modes. For instance, during measurements of the cable mode lifetime in Fig.2{\bf b} in the main text, it is essential to allow the cable modes to evolve freely while preventing the thermalization of the qubits. This is achieved by tuning the qubits into resonance with their readout resonators, which are well thermalized to 50~$\Omega$ loads anchored to the MXC stage through the feedline.
The dissipation process through the readout resonator is illustrated in \rfig{fig_readout_reset}{a}, where qubit $Q_A$ is initially prepared in the $|1\rangle$ state and then tuned down into resonance with its readout resonator $R_A$. The energy rapidly swaps between $Q_A$ and $R_A$ with a period of approximately 3.2~ns, while the amplitude of oscillation decays over time with a characteristic time of about 60~ns, determined by the coupling quality factor of the readout resonator. Within $1~\mathrm{\mu s}$, the qubit reaches a steady state where its residual excitation is determined by its environmental temperature. 
The equilibrium populations of $Q_A$ and $Q_B$ after the dissipation process are shown in \rfigs{fig_readout_reset}{b} and {\bf c}, respectively. For $Q_A$, the excitation populations are reduced to 2.8\%, 5.0\%, 8.5\%, and 11.1\% at $\Thot =$ 1~K, 2~K, 3~K and 4~K, respectively. Similarly, for $Q_B$, the populations are suppressed to 3.3\%, 4.0\%, 6.8\%, and 9.0\% under the same conditions. 
Although these excitation populations are slightly higher than those achieved using the $D$ coupler for qubit reset (see Fig.~3 in the main text and \rfig{fig_iqs}{} for more details), they remain significantly lower than the equilibrium excitation populations without any reset (see \rfig{fig_iqs}{b}). 
These results indicate that the readout resonators provide an effective alternative method for resetting the qubits.

\subsection{Cable benchmark}

\begin{figure}[H]
\centering
\includegraphics[width=0.7\textwidth]{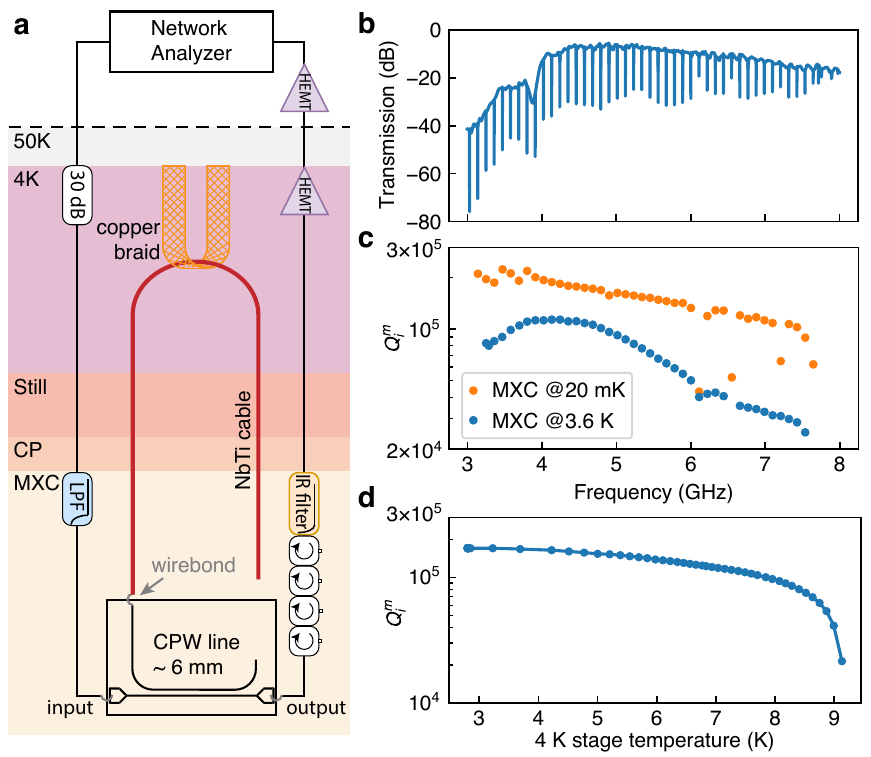}
\caption{
    \label{fig_Qi_VNA}
    {\bf Independent cable test.}
    {\bf a}, Setup for measuring the quality factor of the cable modes.
    The center of the NbTi cable is thermally anchored to the 4~K stage using copper braid.
    {\bf b}, Transmission spectroscopy of the test chip, showing a series of evenly spaced sharp dips, corresponding to the standing wave modes hosted within the cable.
    {\bf c}, Internal quality factors $Q^m_i$ of the cable modes, obtained at the MXC stage temperature of 20~mK and 3.6~K respectively.
    {\bf d}, $Q_i^m$ of the 4.35~GHz mode versus the 4~K plate temperature as the DR warms up.
}
\end{figure}

The NbTi cable used in this experiment, purchased from Keycom Corp., has a superconducting transition temperature of approximately 9.7~K.
In this section, we perform an independent spectroscopy measurement to characterize the quality factor of the cable under various temperatures, as shown in \rfig{fig_Qi_VNA}.
For these tests, the center of the NbTi cable is thermally anchored to the 4~K stage of the DR using copper braid.
One end of the cable is wire-bonded to a dedicated test chip, while the other end is left open.
On the test chip, the cable is connected to a CPW line, which is patterned on a niobium thin film deposited on a sapphire substrate.
The CPW line is approximately 6~mm in length and inductively coupled to a feedline in a hanger geometry before shorted to ground.
The CPW line serves as a quarter-wavelength impedance transformer centered at around 4~GHz~\cite{Niu2023}.
We measure the transmission spectrum of the feedline using a vector network analyzer (VNA).

Before condensing the $^3$He-$^4$He mixture to the DR, the 4~K, still, and MXC stages are maintained at 3.6~K. We perform VNA measurement at this temperature and observe a series of evenly spaced sharp dips in the transmission spectrum (see \rfig{fig_Qi_VNA}{b}), corresponding to the standing wave modes hosted within the cable.
After condensing $^3$He-$^4$He mixture to the DR, the 4~K stage maintains at 3.6~K, but the still stage quickly cools to 0.83~K, and the MXC stage gradually cools to about 20~mK.
We measure the transmission spectrum and extract the internal quality factors~\cite{khalil2012,megrant2012} $Q_i^m$ of the $m$-th cable mode under these two temperature conditions, as shown in \rfig{fig_Qi_VNA}{c}.
The cable modes show relatively high intrinsic quality of $\sim 2\times 10^5$ near 4~GHz, where the 6~mm CPW line minimizes the wirebond loss.
To investigate the cable performance at even higher temperatures, we monitor the 4.35~GHz mode as the DR warms up.
The variation in the quality factor as the 4~K stage temperature increases is shown in \rfig{fig_Qi_VNA}{d}.
Remarkably, the quality factor exhibits no significant degradation until the 4~K stage temperature approaches 8.5~K,  close to the 9.7~K superconducting transition temperature of NbTi.
This observation indicates that the intrinsic loss rate $\kappa_i$ of the cable mode to the thermal environment remains low within the temperature range of interest.
Such robustness is critical for the dynamical reduction of thermal noise in our experiment.
We note that the observed $Q_i^m$ of $\sim 2\times 10^5$ is similar to the NbTi cable quality factors reported in Refs.~\citenum{Kurpiers2017,Burkhart2021,Zhong2021}.

\begin{figure}[H]
    \centering
    \includegraphics[width=0.8\textwidth]{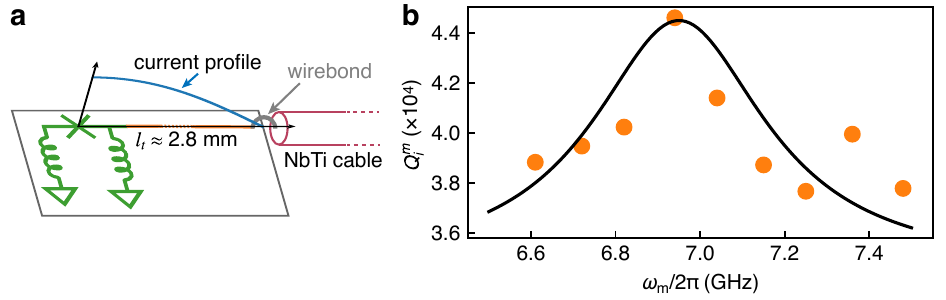}
    \caption{
    \label{fig_Qi_T1r}
    {\bf Quality factors of the standing modes in the communication channel.}
    {\bf a}, Schematic showing the standing mode current profile along the $\lambda/4$ impedance transformer. When frequency matched, the current node coincides with the wirebond joint.
    {\bf b}, The $Q_i^m$ of each mode versus the mode frequency $\omega_m/2\pi$, calculated from the single-photon lifetime. The solid line is fitted using the equation from Ref.~\citenum{Niu2023}.
    }
\end{figure}

Because the thermal occupation decreases with the operating frequency, we choose to operate at higher frequencies, and shorten the $\lambda/4$ CPW line to 2.8~mm to align the standing mode to approximately 7~GHz in Alice and Bob.
In \rfig{fig_Qi_T1r}{}, we present the quality factors derived from the single-photon lifetime measurements similar to that in Fig.2{\bf b} of the main text, with the $D$ coupler turned off.
We see that the mode at around 7~GHz exhibits the highest quality of about $4.4\times 10^4$.
The mode $R_c$ at 7.48~GHz has a slightly lower quality factor, but its frequency is closer to the maximum qubit frequency, therefore we choose this mode for communication in the end.
The quality of the standing modes in the quantum communication setup is much smaller than the results from the cable test \rfig{fig_Qi_VNA}{}, partly because the NbTi cable quality degrades at higher frequencies~\cite{Kurpiers2017,Burkhart2021}.
Also it is more difficult to match both cable-chip joints simultaneously to a single mode, especially at higher operating frequencies.

\section{Quantum thermodynamics}

\subsection{Qubit-channel interaction at various temperatures}

\begin{figure}[H]
\centering
\includegraphics[width=1.0\textwidth]{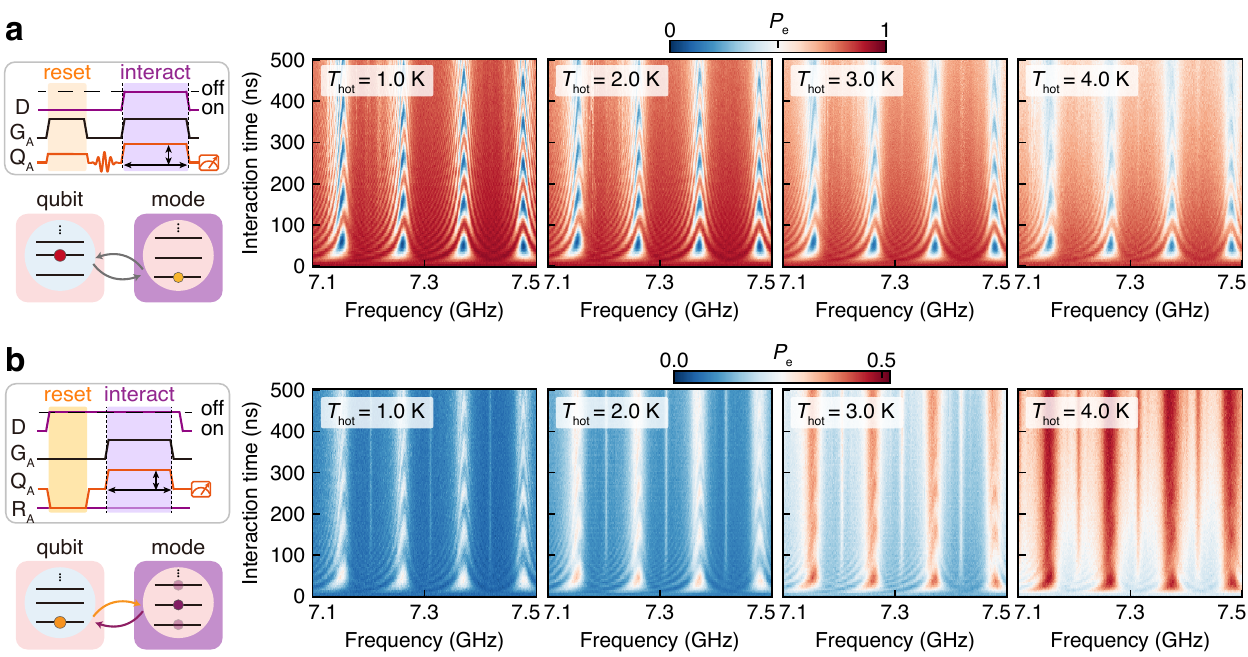}
\caption{\label{rabi_vs_Thot}
    {\bf Interaction between $Q_A$ and cable modes at different $\Thot$.}
    {\bf a},  Vacuum Rabi oscillations between $Q_A$ and the radiatively cooled cable modes at various $\Thot$. The thermal photons in each mode are dynamically reduced by radiative cooling prior to the interaction.
    {\bf b}, Interaction between $Q_A$ and the warm cable modes without prior cooling. We first reset $Q_A$ by tuning into resonance with its readout resonator $R_A$, then use it to detect the thermal photons in the cable modes. 
    % The oscillation chevron fades as $\Thot$ increases, indicating high photon occupation in the mode $m$.
    Left: the experimental control sequence and schematic diagram.
}
\end{figure}

As the decoherence rates remain lower than the interaction rate ($g_n^m/2\pi \approx 5$~MHz) between the qubits and the cable modes, coherent interactions between them are readily observable.
In \rfig{rabi_vs_Thot}{a}, we demonstrate vacuum Rabi oscillations between $Q_A$ and the cable modes. An illustration of the control sequence and experimental diagram is shown on the left. The $D$ coupler is initially turned on to reset both the qubit and cable modes (see \rfig{fig_cool_therm}{} for details). A subsequent $\pi$-pulse prepares $Q_A$ in its first excited state $|1\rangle$. We then vary the qubit frequency to interact with each mode. During the interaction, the $D$ coupler is turned ``off'' to avoid loss of coherent photons. By measuring the excitation probability $P_e$ of $Q_A$, we observe coherent vacuum Rabi oscillations between the qubit and the modes across a $\Thot$ range from 1~K to 4~K. Although faster rethermalization of the modes at $\Thot = 4$~K leads to more rapid decay of the coherent 
oscillations, coherent exchange remains observable up to a few hundred nanoseconds, ensuring reliable quantum state transfer through the channel.

Coherent oscillations are also evident between the qubit and thermal photons within the modes.  In \rfig{rabi_vs_Thot}{b}, we demonstrate the interaction between $Q_A$, initialized in the $|0\rangle$ state, and the cable modes with high thermal occupation. 
To achieve this, we set the $D$ coupler to the ``off'' point, where the standing modes remain warm, while resetting $Q_A$ alone by tuning it into resonance with its readout resonator $R_A$ (see \rfig{fig_readout_reset}{}). Subsequently, we tune up the $G_A$ coupler to a coupling strength of $g^m_A/2\pi\approx5$ MHz, and vary the frequency of $Q_A$ to interact with different standing-wave modes. The thermal photons initially present in the modes are exchanged between the qubit and the modes, resulting in a clearly visible oscillation chevron pattern at $\Thot=1.0~\mathrm{K}$, indicative of a primary thermal population in the $|1\rangle$ Fock state. As $\Thot$ increases, the chevron patterns gradually fade out, reflecting the rapid degradation of the system coherence with increased thermal noise.

\subsection{Cooling and rethermalization at various temperatures}

\begin{figure}[H]
\vspace{-5pt}
\centering
\includegraphics[width=0.8\textwidth]{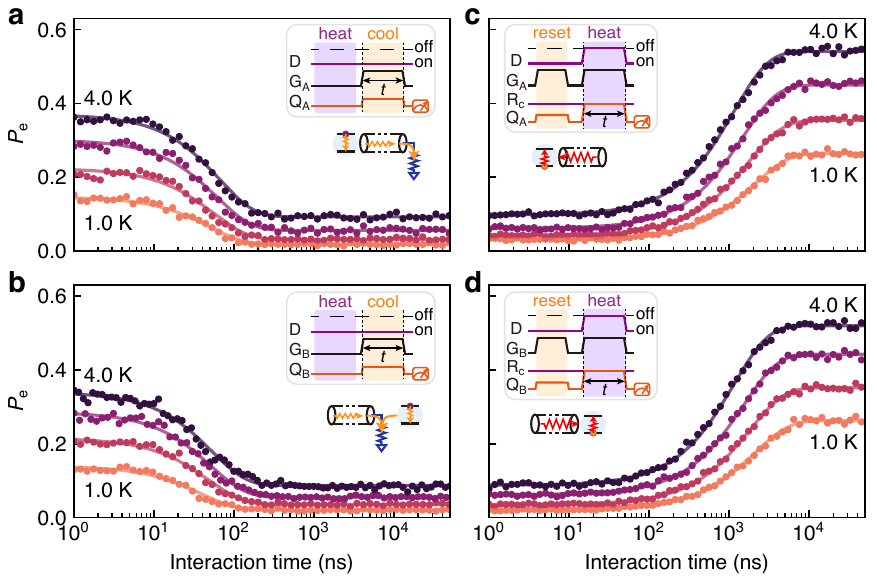}
\vspace{-5pt}
\caption{\label{fig_cool_therm}
    {\bf Cooling and rethermalization at various $\Thot$.}
    {\bf a}, {\bf b}, Cooling of $Q_A$ amd $Q_B$ through the mode $R_c$ at different $\Thot$, with the $D$ coupler suddenly turned to the ``on'' state. The qubits are initially in a thermal state characterized in \rfig{fig_iqs}{b}, where both the gmon coupler and the $D$ coupler are off.
    The solid lines are simple exponential fits with characteristic time of 56~ns, 55~ns, 61~ns, and 65~ns for $Q_A$, and 40~ns, 49~ns, 50~ns, and 52~ns for $Q_B$, at $\Thot=$ 1~K, 2~K, 3~K, 4~K, respectively.
    {\bf c}, {\bf d}, Rethermalization of $Q_A$ and $Q_B$ when coupled to $R_c$ at various $\Thot$, where the $D$ coupler is suddenly turned to the ``off'' state. Both the qubits and $R_c$ are initially reset to their ground states. The qubits are tuned into resonance with $R_c$ while the $D$ coupler is turned off, resulting in thermalization of the qubits to a steady state with a higher excitation probability $P_e$ (see \rfig{fig_iqs}{c}).
    The solid lines represent exponential fits with characteristic time of $\tau = 1.97$~$\mu$s, 1.72~$\mu$s, 1.41~$\mu$s, and 1.14~$\mu$s for $Q_A$, and 2.06~$\mu$s, 1.76~$\mu$s, 1.49~$\mu$s, and 1.15~$\mu$s for $Q_B$, at $\Thot=$ 1~K, 2~K, 3~K, 4~K, respectively.
    Inset shows the control sequences and illustration on the thermal photon flow.
}
\vspace{-5pt}
\end{figure}

In the main text, we discussed the cooling and rethermalization dynamics of the qubit-mode system at $\Thot=4.0~\mathrm{K}$. Here, in \rfig{fig_cool_therm}{}, we extend to various $\Thot$ values, focusing on the mode $R_c$, to evaluate the performance of the $D$ coupler in cooling both the qubit and the thermal channel.
To cool/reset the qubit, we activate the gmon coupler and tune the qubit into resonance with the radiatively cooled mode $R_c$, keeping the $D$ coupler in the ``on'' state throughout the process. 
The time evolution of the qubit excitation probability, $P_e$, is shown in \rfigs{fig_cool_therm}{a} and {\bf b}, revealing the cooling dynamics. The qubits are effectively reset to their ground state, with residual thermal excitation of $P_e =$ 0.016, 0.034, 0.041, and 0.095 for $Q_A$, and $P_e =$ 0.019, 0.035, 0.056, and 0.086 for $Q_B$, corresponding to $\Thot =$ 1~K, 2~K, 3~K and 4~K, respectively.

As long as the $D$ coupler is kept in the ``on'' state, both the qubit and the thermal channel can be effectively cooled through dissipation to the cold bath. However, to enable the coherent transmission of microwave quantum states through the channel, the $D$ coupler must be temporarily turned ``off'' to prevent the loss of coherent photons. During this interval, both the thermal channel and the qubit undergo rethermalization, and the rethermalization rate becomes a critical parameter for achieving efficient quantum state transfer.
To investigate the rethermalization dynamics, we monitor the thermal occupation of the channel after the $D$ coupler is turned ``off''. As illustrated in \rfigs{fig_cool_therm}{c} and {\bf d}, the $D$ coupler is first turned ``on'' to reset both the qubit and the channel effectively. It is then temporarily switched ``off'' while the gmon coupler is activated, allowing the qubit to interact with $R_c$ and reach thermal equilibrium. The rethermalization of the channel is quantified by tracking the time evolution of $P_e$. At equilibrium, the qubits exhibit steady-state excitation probabilities of $P_e =$ 0.26, 0.35, 0.45 and 0.56 for $Q_A$, and $P_e =$ 0.26, 0.35, 0.44, and 0.52 for $Q_B$, at $\Thot =$ 1~K, 2~K, 3~K and 4~K, respectively.
In \rfig{fig_cool_therm}{}, we use simple exponential fitting to characterize the cooling and rethermalization of the system, which shows a typical cooling time of 50~ns, and a typical rethermalization time of $1\sim 2$~$\mu$s. These values serve as a guidance for the quantum state transfer experiments, which are completed within 1/10 of the characteristic rethermalization time. A more careful numerical simulation of the system dynamics based on master equations is given in section~\ref{sec_qutip_simu}.

% \subsection{Qubit states at thermal equilibrium.}

\begin{figure}[H]
\centering
\includegraphics[width=0.98\textwidth]{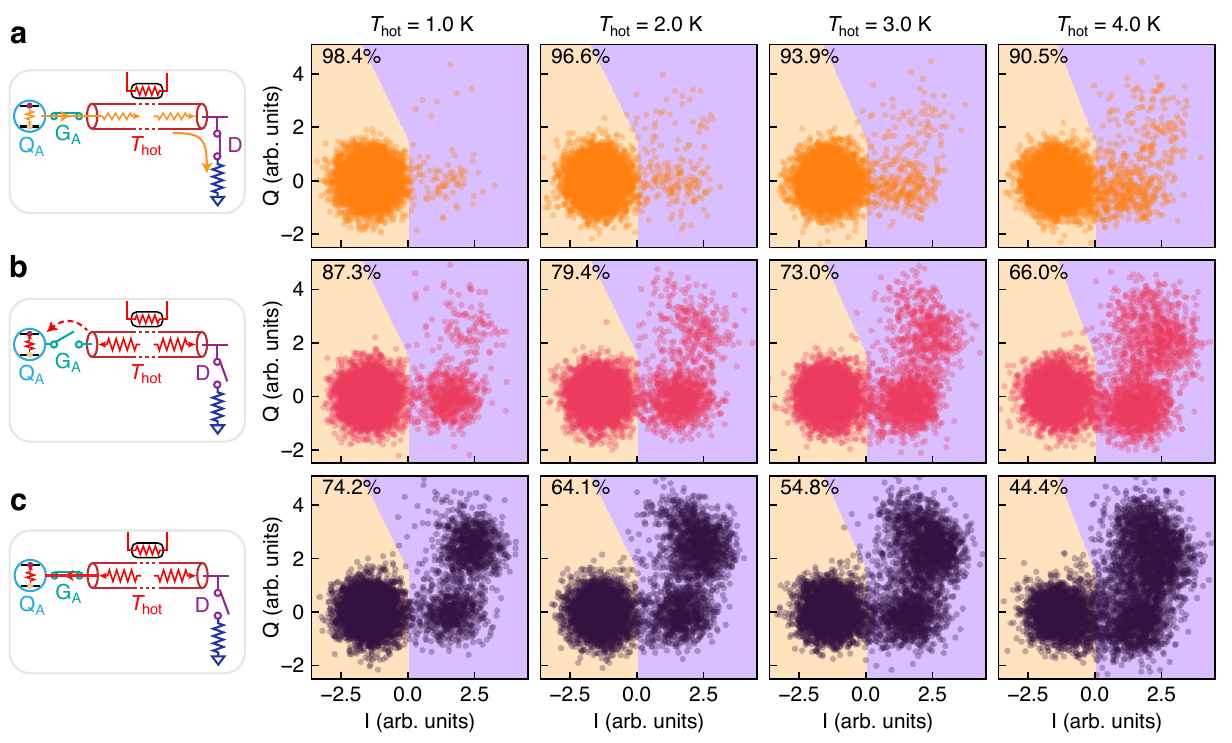}
\caption{\label{fig_iqs}
    {\bf Qubit steady states under different configurations.}
    Single-shot dispersive readout results in the quadrature (IQ) space of $Q_A$ after reaching equilibrium with {\bf a} both the $G_A$ and $D$ couplers are turned on; {\bf b} both the $G_A$ and $D$ couplers are turned off; {\bf c} $G_A$ is turned on while $D$ is turned off.
    The left illustrates the thermal noise flow in these cases. 
    Columns show data at different $\Thot$.
    Annotations indicate the proportion of results discriminating the qubit in the $|0\rangle$ state.
}
\end{figure}

Figure~\ref{fig_iqs} provides further details about the steady states of $Q_A$. The single-shot dispersive readout results of $Q_A$ are presented in the quadrature (IQ) space, obtained after 6000 repetitions.
In \rfig{fig_iqs}{a} we present the readout results of steady state with both $G_A$ and $D$ couplers turned on, corresponding to the final state after the cooling process in \rfig{fig_cool_therm}{a}. The ground state population $P_0$ is annotated in the upper left corner of each panel. Thermal excitations are largely suppressed for all $\Thot$, resulting in $P_e<0.1$ even at 4~K.

In \rfig{fig_iqs}{b}, the steady state of $Q_A$ is displayed with both $G_A$ and $D$ couplers turned off. This corresponds to the initial state of the cooling process in \rfig{fig_cool_therm}{a}, where the qubit should be decoupled from the warm channel ideally. However, significant thermal excitations are observed, with $P_e$ increasing from 0.13 at $\Thot=$ 1~K to 0.34 at $\Thot=$ 4~K. Signatures of $|2\rangle$ or higher-level occupancy also become evident. These results suggest that thermal noise from the warm channel cannot be fully isolated by the gmon coupler. Further improvements, such as implementing bandpass filters or quasiparticle suppression techniques, may enhance the system's thermal isolation and overall performance.

In \rfig{fig_iqs}{c}, we demonstrate the impact of thermal noise when the $D$ coupler is turned ``off'', while $G_A$ remains switched on.
 This corresponds to the final state of the rethermalization process in \rfig{fig_cool_therm}{c}.
In this configuration, thermal excitation $P_e$ exceeds 0.3 at $\Thot=1.0~\mathrm{K}$, and even surpasses 0.5 at $\Thot=4.0~\mathrm{K}$, making reliable qubit state preparation infeasible.

\subsection{Numerical simulation}\label{sec_qutip_simu}
\begin{figure}[H]
\centering
\includegraphics[width=0.5\textwidth]{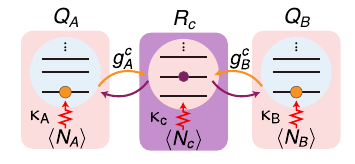}
\caption{\label{fig_model_illustration}
    {\bf Quantum thermodynamics model for the system.}
    Two qubits are coupled to the communication mode $R_c$. Each component is also coupled to an effective heat bath characterized by the coupling rate $\kappa_j$ and thermal occupancy $\langle N_j \rangle$ ($j=A$, $B$, $c$).
}
\end{figure}

Modeling the system dynamics with multiple bosonic modes, which are occupied with a large number of thermal photons, is complicated.
Since the qubit-mode coupling $g_n^m \ll \omega_{\rm{FSR}}$ in the quantum communication experiment, we can ignore irrelevant modes and restrict Eq.~(1) in the main text to a single bosonic mode $R_c$:
\begin{equation}\label{eq_hami}
    H/\hbar=
    \sum_{n=A,B}(
    \omega_n \sigma_n^{\dagger } \sigma_n 
    +\frac{\eta}{2} \sigma_n^{\dagger } \sigma_n^{\dagger } \sigma_n \sigma_n
    ) 
    + \omega_{c} a_{c}^{\dagger } a_{c} 
    +\sum_{n=A,B} g_n^c(
    \sigma_n^\dagger a_c + \sigma_n a_c^\dagger
    ) ,
\end{equation}
where $\sigma_n$ and $a_{c}$ represent the annihilation operators for qubit $Q_n$ and the cable mode $R_c$ at 7.48~GHz, respectively. The qubits are modeled as Duffing oscillators with anharmonicity $\eta$. To account for system decoherence in the presence of a thermal environment, we analyze the dynamics of the system's density matrix $\rho$ using the Lindblad master equation under the Born-Markov approximation:
\begin{equation}\label{eq_master}
    \dot{\rho }( t) =\frac{1}{i\hbar }[ H,\rho ( t)] +\sum_{j}\frac{1}{2}\left[ 2L_{j} \rho ( t) L_{j}^{\dagger } -\rho ( t) L_{j}^{\dagger } L_{j} -L_{j}^{\dagger } L_{j} \rho ( t)\right] ,
\end{equation}
with the corresponding Lindblad operators $L_j$~\cite{schuetz2019}:
\begin{align}
    L_{\phi ,n}       & =\sqrt{\Gamma_{\phi } /2}\left( \sigma_n^{\dagger } \sigma_n\right) , \\
    L_{\downarrow ,n} & =\sqrt{\kappa_n(\langle N_n\rangle +1)} \sigma_n ,                    \\
    L_{\downarrow ,c} & =\sqrt{\kappa_{c}(\langle N_{c}\rangle +1)} a_c ,                     \\
    L_{\uparrow ,n}   & =\sqrt{\kappa_{n}\langle N_{n}\rangle } \sigma_n^\dagger ,            \\
    L_{\uparrow ,c}   & =\sqrt{\kappa_{c}\langle N_{c}\rangle } a_c^\dagger.
\end{align}
The first term represents the pure dephasing of the qubits; the subsequent terms describe the relaxation of each component of the system toward a thermal state with an effective rate $\sim \kappa_j (\langle N_j\rangle +1)$ ($j=A$, $B$, $c$). Here, $\kappa_j $ denotes the coupling rate between the system and its thermal bath, and $\langle N_j\rangle $ is the average thermal occupation in the bath, as illustrated in \rfig{fig_model_illustration}{}.

The effective environment of the system is a composite of all thermal baths coupled to its components. For example, the communication mode $R_{c}$ interacts with two baths: the thermal environment characterized by an average photon number $\langle N_{e}\rangle$ and coupling rate $\kappa_{i}$, and the $50~\Omega$ cold load with an average photon number $\langle N_{\mathrm{cold}}\rangle$ and coupling rate $\kappa_{D}$.
The combined effect of these baths can be equivalently described by an effective environment with the following parameters:
\begin{align}
    \langle N_{c}\rangle   & =\frac{\kappa_{i}\langle N_{e}\rangle +\kappa_{D}\langle N_{\mathrm{cold}}\rangle }{\kappa_{i} +\kappa_{D}} , \\
    \kappa_{c} & =\kappa_{i} +\kappa_{D} ,
\end{align}
which simplifies the analysis of the system's thermal dynamics.

\begin{figure}[H]
  \centering
  \includegraphics[width=0.96\textwidth]{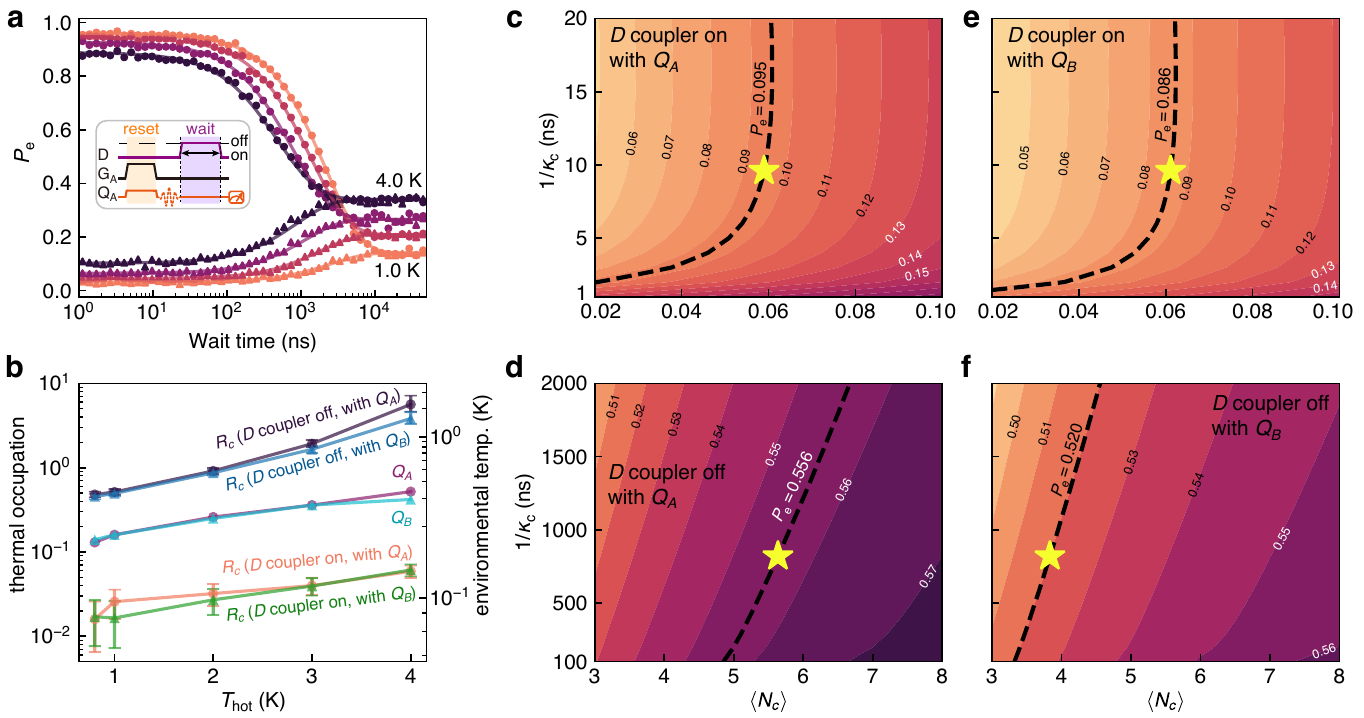}
  \caption{
  \label{fig_qutip_simu} 
  % {\bf Estimation of temperature with numerical simulation.}
  {\bf Estimation of effective temperature and thermal photon occupancy.}
  {\bf a}, Evolution of $|1\rangle$ state relaxation (circle dots) and 
$|0\rangle$ state thermalization (triangular dots) of $Q_A$ at various $\Thot$ from 1~K to 4~K. Solid lines are master equation simulation results. Inset shows the control sequence.
  {\bf b}, The thermal photon occupancy, as well as the effective environmental temperature for both qubits, and $R_c$ with $D$ coupler on and off, respectively.
  Data for the qubits are from master equation simulations in {\bf a}.
  Data for mode $R_c$ are estimated from steady state analysis using either $Q_A$ (circle dots) or $Q_B$ (triangular dots) shown in ({\bf d}--{\bf f}).
  Errorbars for $R_c$ are estimated assuming a $\pm 0.01$ variation in the qubit steady state $P_e$.
  {\bf c}, {\bf d}, Numerical steady state $P_e$ of $Q_A$ when coupled to $R_c$, as a function of $\langle N_c\rangle$ and $\kappa_c$ at $\Thot=4.0~\mathrm{K}$. 
   $\kappa_A$ and $\langle N_A \rangle$ are taken from the fitting results in {\bf a}.
  The dashed lines are contours for the experimentally observed value of $P_e=0.095$ and 0.556 respectively. 
  The stars indicate the points where $\kappa_c$ matches the value measured at $\Thot=0.83~\mathrm{K}$.
  {\bf e}, {\bf f}, Steady state $P_e$ of $Q_B$ coupled to $R_c$ at $\Thot=4$~K, similar to {\bf c} and {\bf d}.
  }
\end{figure}

We numerically simulate the system dynamics described by Eq.~\ref{eq_hami} and Eq.~\ref{eq_master} using QuTiP~\cite{johansson2012}. By fitting the simulation results to experimental data, we estimate system parameters, including the number of thermal photons in the environment $\langle N_j \rangle$ and the system-environment coupling rate $\kappa_j$.
In \rfig{fig_qutip_simu}{a}, we compare the numerical simulations with $|1\rangle$ state relaxation and 
$|0\rangle$ state thermalization of $Q_A$ at various $\Thot$.
Importantly, the simulations do not assume a perfect initial state ($|0\rangle$ or $|1\rangle$) but instead start with a mixed state characterized by the initial $P_e$, which is set to the experimentally measured value at the beginning of the evolution.
Focusing solely on $Q_A$ ($Q_B$) while neglecting contributions from other components, the fitted parameters for $Q_A$ ($Q_B$) are summarized in Table~\ref{tab_qb_kappa}, and the thermal occupancy is plotted in \rfig{fig_qutip_simu}{b}. The observed decrease in $1/\kappa_A$ ($1/\kappa_B$) with $\Thot$ suggests that the the qubit coherence degrades as the thermal noise level increases.
The effective temperatures are calculated from the thermal occupancy according to the Bose-Einstein distribution.

\begin{table}[H]
\centering
\begin{tabular}{
    |
    >{\centering\arraybackslash}p{5.0cm}
    |
    >{\centering\arraybackslash}p{1.5cm}
    >{\centering\arraybackslash}p{1.5cm}
    >{\centering\arraybackslash}p{1.5cm}
    >{\centering\arraybackslash}p{1.5cm}
    >{\centering\arraybackslash}p{1.5cm}
    |
}
    \hline
    \hline
    $\Thot$               & 0.83~K                & 1.0~K                 & 2.0~K                 & 3.0~K                 & 4.0~K                 \\
    \hline
    $1/\kappa_A$          & 2.74~$\mathrm{\mu s}$ & 2.80~$\mathrm{\mu s}$ & 2.05~$\mathrm{\mu s}$ & 1.37~$\mathrm{\mu s}$ & 1.08~$\mathrm{\mu s}$ \\
    $\langle N_A \rangle$ & 0.13                  & 0.16                  & 0.26                  & 0.36                  & 0.52                  \\
    \hline
    $1/\kappa_B$          & 3.54~$\mathrm{\mu s}$ & 3.48~$\mathrm{\mu s}$ & 2.75~$\mathrm{\mu s}$ & 1.83~$\mathrm{\mu s}$ & 1.30~$\mathrm{\mu s}$ \\
    $\langle N_B \rangle$ & 0.14                  & 0.16                  & 0.25                  & 0.36                  & 0.42                  \\
    \hline
    $\langle N_c \rangle$ ($D$ coupler on, with $Q_A$) & 0.016    & 0.026    & 0.032    & 0.040    & 0.059    \\
    $\langle N_c \rangle$ ($D$ coupler on, with $Q_B$) & 0.017    & 0.017    & 0.027    & 0.039    & 0.061    \\
    $\langle N_c \rangle$ ($D$ coupler off, with $Q_A$) & 0.48    & 0.52    & 0.92    & 1.92    & 5.64    \\
    $\langle N_c \rangle$ ($D$ coupler off, with $Q_B$) & 0.46    & 0.50    & 0.87    & 1.65    & 3.83    \\
    \hline
    \hline
\end{tabular}
\caption{\label{tab_qb_kappa}
    {\bf Thermodynamic parameters of the system.}
    $\kappa_n$ and $\langle N_n \rangle$ ($n=A, B$) are fitted parameters obtained from the numerical simulations as shown in \rfig{fig_qutip_simu}{a}.
    The thermal occupancy $\langle N_c \rangle$ for $R_c$ is estimated from the steady state analysis using $Q_A$ or $Q_B$, as shown in in \rfigs{fig_qutip_simu}{c} to {\bf f}.
}
\end{table}

In addition to analyzing the qubits alone, it is crucial to estimate the thermal occupation of the cable mode $R_c$, with or without radiative cooling.
As mentioned in the main text, superconducting qubits based microwave thermometry has been demonstrated at sub-Kelvin temperatures~\cite{Scigliuzzo2020,Wang2021,Lvov2024}, but extending these methods to liquid helium temperature is quite challenging.
We note that the simulation of the qubit dynamics in \rfig{fig_qutip_simu}{a} is valid because the thermal occupancy of the qubit alone is small (see Table~\ref{tab_qb_kappa}).
With many thermal photons in the channel, simulating the system dynamics becomes infeasible, as the state preparation and measurement processes are spoiled by thermal noise.
Here we instead adopt a steady state approach, which is done by comparing the system's steady state in numerical simulation with the experimentally measured steady state $P_e$ from \rfig{fig_iqs}{}, see \rfigs{fig_qutip_simu}{c} to {\bf f}.
The steady state analysis is robust against errors introduced by qubit decay and rethermalization during the measurement process.
Furthermore, the steady state $P_e$ is predominantly determined by  the thermal occupancy $\langle N_j \rangle$ and the system-environment coupling rate $\kappa_j$ ($j=A$, $B$, $c$), regardless of the system initial state. 
As shown in \rfigs{fig_qutip_simu}{c} to {\bf f}, we observe a clear dependence of the steady state $P_e$ on $\kappa_c$ and $\langle N_c \rangle$.
\rfigs{fig_qutip_simu}{c} and {\bf d} highlight the steady state $P_e$ of $Q_A$, corresponding to the configuration where the $D$ coupler is turned ``on''and ``off'' respectively.
The black dashed lines in \rfigs{fig_qutip_simu}{c} and {\bf d} are contours for $P_e=0.095$ and 0.556, corresponding to the experimentally measured steady state excitation of $Q_A$ at $\Thot=4.0~\mathrm{K}$ with the $D$ coupler is ``on'' (\rfig{fig_iqs}{a}) and ``off'' (\rfig{fig_iqs}{c}), respectively.
The yellow stars in \rfigs{fig_qutip_simu}{c} and {\bf d} indicate the probable parameters based on the mode coherence measured at $\Thot=0.83~\mathrm{K}$ (see Fig.~2 in the main text).
The variation, along with the estimated values marked by the yellow stars, is plotted in \rfig{fig_qutip_simu}{b} for various $\Thot$.
At $\Thot=4.0~\mathrm{K}$, we estimate {\small $\langle N_c \rangle \approx 5.64$} when the $D$ coupler is turned ``off''. Conversely, when the $D$ coupler is turned ``on'', we estimate $\langle N_c \rangle \approx 0.06$.
Similar steady state analysis of $R_c$, but instead using the steady state excitation of $Q_B$ for estimation, is shown in \rfigs{fig_qutip_simu}{e} and {\bf f}.
We again confirm $\langle N_c \rangle \approx 0.06$ when the $D$ coupler is turned ``on'', but there is observable discrepancy for the thermal occupancy when the $D$ coupler is turned ``off'', estimated using different qubits, suggesting this method is also limited when there is large thermal photons in the system.
If we take an average of the results from both qubits, we have $\langle N_c \rangle \approx 4.73$.
This corresponds to a thermal photon reduction ratio of 79, very close to the estimation of 86 in the main text using $\kappa_i$ and $\kappa_D$ measured at $\Thot=0.83$~K.

\section{State transfer and remote entanglement at other temperatures}

\begin{figure}[H]
  \centering
  \includegraphics[width=0.98\textwidth]{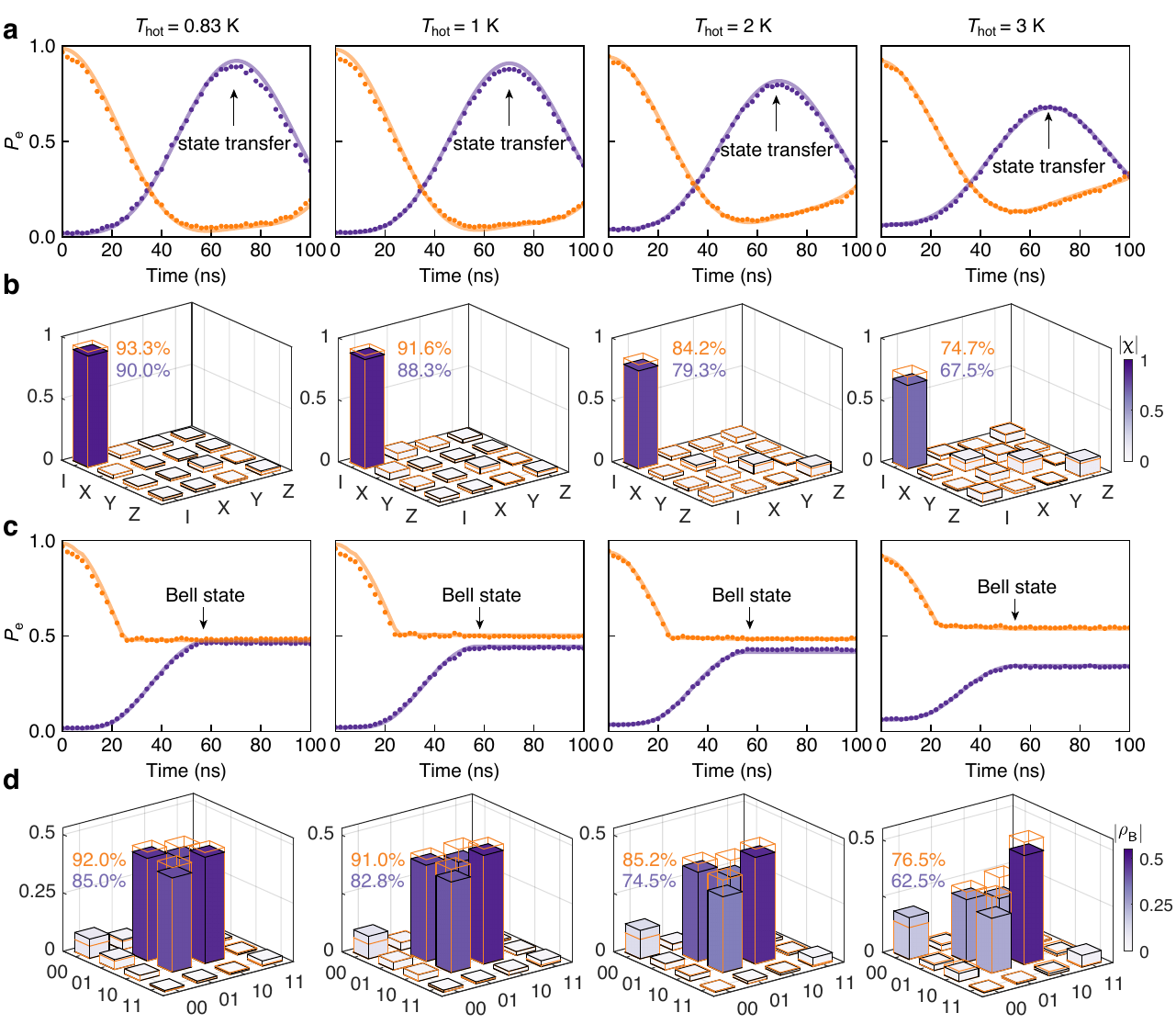}
  \caption{\label{bell_vs_temperature} 
  {\bf Quantum state transfer and remote entanglement generation across various $\Thot$.}
  {\bf a}, Transferring a photon from $Q_A$ to $Q_B$ via $R_c$. Orange (purple) circles represent the excited state populations of $Q_A$ ($Q_B$).
  {\bf b}, Process matrices ($\chi$) characterizing the quantum state transfer at $\Thot = 0.83~\mathrm{K}$, $1~\mathrm{K}$, $2~\mathrm{K}$, and $3~\mathrm{K}$. Process fidelities ($\mathcal{F}_\chi$) are shown for cases without SPAM error correction (purple bars) and with SPAM error correction (orange frames).
  {\bf c}, Transferring half a photon from $Q_A$ to $Q_B$.
  {\bf d}, Density matrices ($\rho_B$) for the Bell state generated at $t=53$~ns. The corresponding fidelities ($\mathcal{F}_\rho$) are presented for cases without (purple bars) and with (orange frames) SPAM error correction.
  }
\end{figure}

In Fig.~4 of the main text, we show the quantum state transfer and remote entanglement generation through the thermal channel at $\Thot=4$~K. \rfig{bell_vs_temperature}{} extends these results to a range of channel temperatures, $\Thot = 0.83~\mathrm{K}$, $1~\mathrm{K}$, $2~\mathrm{K}$, and $3~\mathrm{K}$, respectively.
In \rfig{bell_vs_temperature}{a}, we show the transfer of a single photon from $Q_A$ to $Q_B$ via mode $R_c$. This process follows the same pulse sequence described in Fig. 4 of the main text, where the $D$ coupler is turned ``off'', and both $G_A$ and $G_B$ are activated with matching coupling strengths ($g_A^c = g_B^c \approx 2\pi\times 5~\mathrm{MHz}$).
After an evolution time of $t = 65~\mathrm{ns}$, the photon population in $Q_B$ reaches its maximum. The efficiency of quantum state transfer is characterized using quantum process tomography~\cite{Zhong2021}, with the reconstructed process matrices $\chi$ displayed in \rfig{bell_vs_temperature}{b}. 
The corresponding process fidelities, $\mathcal{F}_\chi$, are determined to be 90.0\%, 88.3\%, 79.3\%, and 67.5\% for $\Thot = 0.83~\mathrm{K}$, $1~\mathrm{K}$, $2~\mathrm{K}$, and $3~\mathrm{K}$, respectively. These fidelities are calculated directly from the raw data, while the values highlighted by orange frames in \rfig{bell_vs_temperature}{b} represent the corrected fidelities after applying the SPAM calibration.

Figure \ref{bell_vs_temperature}{\bf c} presents the generation of remote entanglement between $Q_A$ and $Q_B$. In this process, half a photon is first transferred from $Q_A$ to mode $R_c$, which subsequently relays the state to $Q_B$.
The Bell state fidelities, $\mathcal{F}_\rho = \langle \psi_B | \rho_B | \psi_B \rangle$, derived from raw data are 85.0\%, 82.8\%, 74.5\%, and 62.5\% for $\Thot = 0.83~\mathrm{K}$, $1~\mathrm{K}$, $2~\mathrm{K}$, and $3~\mathrm{K}$, respectively. After applying SPAM correction, the fidelities improve to 92.0\%, 91.0\%, 85.2\%, and 76.5\%, as shown in \rfig{bell_vs_temperature}{d}.

The solid lines in \rfigs{bell_vs_temperature}{b} and {\bf d} represent the results of QuTiP simulations. The initial states for these simulations include thermal excitations consistent with conditions when the qubit is cooled using the $D$ coupler. For each $\Thot$, system parameters $\kappa_i$ and thermal photon numbers $\langle N_j\rangle$ ($j=A, B, c$) and the coupling rates $g_n^c$ ($n=A,B$) are fitted around the values calibrated at $\Thot=0.83~\mathrm{K}$ to match the experimental data. The simulated process fidelities $\mathcal{F}_\chi$ and Bell state fidelities $\mathcal{F}_\rho$ in Figs.~4{\bf d} and {\bf h} are obtained using the same initial states and parameters.

%\clearpage

\bibliographystyle{naturemag}
\bibliography{ref}